\documentclass[11pt,a4paper]{article}
\usepackage{amsmath}
\usepackage{bbm}
\usepackage{amssymb}
\usepackage{amscd}
\usepackage{graphicx}
\usepackage{caption}
\usepackage{chngcntr}
\usepackage{textcomp}
\usepackage{float}
\usepackage{subfig}
\usepackage{multirow}
\usepackage{booktabs}
\usepackage{array}
\usepackage[colorlinks,citecolor=blue,urlcolor=blue]{hyperref}
\usepackage{color}
\usepackage{amsthm}
\usepackage{enumerate}
\usepackage[round]{natbib}
\usepackage{bm}
\usepackage{algorithm}
\usepackage{algpseudocode}

\newtheorem{theorem}{Theorem}
\newtheorem{proposition}{Proposition}
\newtheorem{remark}{Remark}

\newtheorem{assumption}{Assumption}

\newtheorem{example}{Example}

\title{Generalized linear models under distribution uncertainty}
\author{Xifeng Li\thanks{Corresponding author. School of Mathematics, Shandong University, PR China, lixfay@mail.sdu.edu.cn.}\quad and \quad Shuzhen Yang\thanks{Shandong University-Zhong Tai Securities Institute for Financial Studies, Shandong University, PR China, yangsz@sdu.edu.cn.}}

\date{}
\begin{document}
\maketitle
\begin{abstract}
In this paper, we study generalized linear models under distribution uncertainty, where the random intercept has mean uncertainty and takes values in an unknown bounded interval. We develop a novel two-step method, named Firth-corrected Expectation-Maximization with Moving Block (FEMMB), to estimate the regression coefficients as well as the lower and upper bounds of the intercept. A Firth bias correction is incorporated into the EM algorithm to mitigate the incidental parameters problem, which is prominent in generalized linear models with many nuisance intercepts. Under mild regularity conditions, we establish the consistency and asymptotic normality of the estimated regression coefficients, and the consistency of the estimated intercept bounds. Simulation studies and a real-data application to an imbalanced credit card dataset demonstrate the effectiveness of the proposed method.\\

\end{abstract}

\section{Introduction}
Generalized linear models (GLMs) extend linear regression to accommodate non-normal responses, such as binary, count, and nonnegative continuous data \citep{mccullagh1989generalized}. Standard GLMs assume a constant intercept across all observations, but this assumption is often too restrictive, as the intercept can exhibit substantial heterogeneity \citep{ji2023imbalanced, fan2024environment, knight2026fast}. Ignoring such heterogeneity can distort coefficient estimates and impair prediction accuracy. Standard fixed- and random-effects models are commonly used to deal with intercept heterogeneity when group structures are already known. In many practical scenarios, however, the intercept may vary across latent subgroups or time segments, rendering the standard approaches inapplicable.

Building on the preliminary work of \cite{peng2004filtration,peng2005nonlinear}, \cite{peng2007g} introduced the theory of sublinear expectation to address distribution uncertainty. Specifically, the theory considers a family of probability measures $\{P_{\theta}\}_{\theta\in\Theta}$ rather than a single probability measure $P_{\theta_0}$, which naturally captures the uncertainty of a random variable. For a comprehensive exposition, we refer readers to Peng's plenary lecture at the 2010 International Congress of Mathematicians \citep{PengICM2010} and his monograph \citep{peng2019nonlinear}. Inspired by the theory of sublinear expectation, we study GLMs under distribution uncertainty, where the random intercept $\nu$ has mean uncertainty and takes values in an unknown bounded interval $[\underline{\beta}_0,\overline{\beta}_0]$. Our goal is to estimate the regression coefficients as well as the lower and upper bounds of the intercept.

\cite{li2026robust} considered linear regression case and developed a two-step method named Expectation-Maximization with Moving Block (EMMB) to estimate the model parameters. They first use an EM algorithm to estimate the regression coefficients under an unknown group structure, treating the intercepts as missing data within small windows of size $n_0$. Then, a moving block procedure is used to estimate the lower and upper bounds of the intercept. However, when extended to generalized linear models such as logistic regression, their EM algorithm suffers from the well-known incidental parameters problem \citep{neyman1948consistent}.

To mitigate this problem, we incorporate a Firth bias correction \citep{firth1993bias, kosmidis2009bias, kosmidis2021jeffreys} into the EM algorithm. Unlike the panel jackknife and analytic bias correction of \cite{hahn2004jackknife}, which reduce bias by correcting the resulting coefficient estimator, our correction is applied directly to the log-likelihood function. This is natural because the incidental parameter bias originates primarily from the estimation of window-specific intercepts. We adopt the same moving block procedure as \cite{li2026robust}, and name our method Firth-corrected Expectation-Maximization with Moving Block (FEMMB). In the first step, a Firth-corrected EM algorithm is employed to obtain a bias-reduced estimator of $\bm{\beta}$. In the second step, a moving block procedure is applied to estimate the upper and lower bounds $\overline{\beta}_0$ and $\underline{\beta}_0$. Under mild regularity conditions, we establish the consistency and asymptotic normality of the Firth-corrected EM estimator $\hat{\bm{\beta}}_{\mathrm{FEM}}$, as well as the consistency of $\hat{\overline{\beta}}_0$ and $\hat{\underline{\beta}}_0$ obtained by the moving block procedure.

Comprehensive simulation studies demonstrate the effectiveness of the proposed FEMMB method. For logistic regression, where EMMB suffers from the severe incidental-parameter problem, FEMMB substantially reduces the estimation bias and MSE. For Poisson regression, where the EMMB estimator is already consistent, FEMMB coincides with EMMB. Both methods substantially outperform pooled MLE. These results confirm that modeling the intercept heterogeneity through a piecewise constant structure is inherently valuable, and that the Firth correction is effective when needed while incurring no additional costs otherwise. The practical utility of the FEMMB method is further illustrated through a real-data application to credit card default prediction. Our proposed FEMMB method outperforms all competing methods in both F1-score and balanced accuracy, achieving a better trade-off between identifying genuine defaulters and controlling misclassification of normal clients.

The rest of the paper is organized as follows. Section \ref{sec:2} introduces generalized linear models under distribution uncertainty. Section \ref{sec:3}
develops the FEMMB method for estimating the unknown parameters $\bm{\beta}, \underline{\beta}_0$ and $\overline{\beta}_0$. Section \ref{sec:4} establishes the theoretical properties of the obtained estimators. Section \ref{sec:5} presents simulation studies to evaluate the finite-sample performance of the FEMMB method. Section \ref{sec:6} provides a real-data application. Finally, we conclude the paper with some discussions. Technical proofs are relegated to the Appendix.

\section{Background and model setup}\label{sec:2}

\subsection{Preliminaries on generalized linear models}
A generalized linear model(GLM) describes a relationship between the mean of the response $y \in \mathbb{R}$ and the vector of predictors $\mathbf{x}=(x_1,\ldots,x_p)^{\top} \in \mathbb{R}^p$. Assume that the response $y$ has a density belongs to the exponential family with the form
\begin{equation}
f(y; \theta, \phi) = \exp\left(\frac{y\theta - b(\theta)}{a(\phi)} + c(y, \phi)\right),
\end{equation}
where $a(\cdot), b(\cdot)$ and $c(\cdot)$ are some specific functions, with $a(\phi)>0$. The parameter $\theta$ is called the canonical parameter and $\phi$ is the dispersion parameter. Usually, we are more concerned with the parameter $\theta$ rather than $\phi$, so $\phi$ is also referred to as a nuisance parameter, akin to the variance $\sigma^{2}$ in linear regression model with Gaussian noise. For a one-parameter exponential family $a(\phi)=1$.

Besides, under some regularity conditions, based on the properties of the first and second derivatives of the log-likelihood function, we can obtain
$$
\mu = \mathbb{E}[y]=b^{\prime}(\theta)
$$

$$
\mathrm{Var}[y]=V(\mu)a(\phi)=b^{\prime\prime}(\theta)a(\phi)
$$
It is obvious that the expectation of $y$ depends solely on $\theta$, while the variance of $y$ depends on the parameter of interest $\theta$ and the nuisance parameter $\phi$.

The link function $g(\mu) = \eta$ connects the conditional expectation $\mu = \mathbb{E}[y \mid \mathbf{x}]$ and the linear predictor $\eta$. Here $g(\cdot)$ is a monotonic differentiable function and the linear predictor $\eta$ takes the form
\begin{equation}
\eta = \mathbf{x}^{\top}\bm{\beta} + \beta_0,
\end{equation}
where $\bm{\beta}=(\beta_1,\ldots,\beta_p)^{\top} \in \mathbb{R}^p$ is the vector of regression coefficients and $\beta_0 \in \mathbb{R}$ is the intercept term.

Different generalized linear models have different link functions. For example, in linear regression model with Gaussian noise, the response $y$ is continuous and $g(\mu) = \mu$; in logistic regression model, we have a binary $y$ and $g(\mu) = \ln \left( \frac{\mu}{1-\mu} \right)$; in Probit regression model, $y$ is also binary but $g(\mu) = \Phi^{-1}(\mu)$; and in Poisson regression model, $y$ is a nonnegative integer and $g(\mu) = \ln(\mu)$. In the case that $\theta=\eta$, the link function is called canonical link function. Except for the link function of the Probit regression model, the other three link functions mentioned above are all canonical.

\subsection{Generalized linear models with distribution uncertainty}\label{sec:GGLM}

Let $\xi$ be a random variable with distribution uncertainty in the sense that its possible distributions form a family $\{F_{\theta}\}_{\theta\in\Theta}$. Suppose that $\mathbb{E}_{F_{\theta}}[|\xi|]<\infty$ for all $\theta \in \Theta$. The upper expectation of $\xi$ is defined as

\begin{equation}\label{eq:upper}
\hat{\mathbb{E}}[\xi]:=\sup_{\theta \in \Theta} \mathbb{E}_{F_{\theta}}[\xi],
\end{equation}
where $\mathbb{E}_{F_{\theta}}[\cdot]$ is the classical expectation under distribution $F_{\theta}$.

This upper expectation represents the maximum expectation of $\xi$ over all possible distributions in $\{F_{\theta}\}_{\theta\in\Theta}$ and constitutes a special case of sublinear expectation. Here, we do not conduct our analysis directly in the sublinear expectation framework of \cite{peng2019nonlinear}, but only adopt the above definition. Other settings in this paper, such as independence and the independent and identically distributed (i.i.d.) assumption, are understood in the classical probabilistic sense. Throughout the paper, $\mathbb{E}_{F_{\theta}}[\cdot]$ denotes the classical expectation under distribution $F_\theta$, while $P$ and $\mathbb{E}$ denote classical probability and expectation under the true underlying distribution, and $\hat{\mathbb{E}}[\cdot]$ denotes the upper expectation given in \eqref{eq:upper}.

In this paper, we investigate the generalized linear models under distribution uncertainty. To be specific, we no longer assume that the intercept of the linear predictor $\eta$ is a constant $\beta_0$, but rather assume that
\begin{equation}
\eta = \mathbf{x}^{\top}\bm{\beta} + \nu,
\end{equation}
where $\nu$ is an unobserved random variable taking values in $[\underline{\beta}_0,\overline{\beta}_0]$ and has distribution uncertainty. That is, its possible distributions form a family $\{F_{\theta}\}_{\theta\in\Theta}$, which includes the uniform distribution $U(\underline{\beta}_0,\overline{\beta}_0)$, the degenerate distributions $\delta_{\underline{\beta}_0}$ and $\delta_{\overline{\beta}_0}$, among others. By definition of the upper expectation, we have
$$
\overline{\beta}_0=\hat{\mathbb{E}}[\nu]\qquad\text{ and }\qquad\underline{\beta}_0=-\hat{\mathbb{E}}[-\nu].
$$
Moreover, the vector of predictors $\mathbf{x}$ follows a fixed distribution $F_{\mathbf{x}}(\cdot)$, and $\nu$ is independent of $\mathbf{x}$. The link function $g$ satisfies

\begin{equation}
g\left(\mathbb{E}[y\mid \mathbf{x},\nu]\right)=\eta.
\end{equation}

To quantify the conditional mean under distribution uncertainty, we introduce the following conditional upper expectation
$$
\hat{\mathbb{E}}[y\mid \mathbf{x}]=\sup_{\theta \in \Theta} \mathbb{E}_{F_{\theta}}[y\mid \mathbf{x}].
$$
Based on this definition, we have the following proposition for the generalized linear regression models with distribution uncertainty.

\begin{proposition}\label{prop:glm-interval}
For the generalized linear model with distribution uncertainty, the uncertainty of the conditional mean $\mathbb{E}[y\mid \mathbf{x}]$ can be characterized by the interval $[\underline{\mu},\overline{\mu}]$, where
$$
\underline{\mu}=-\hat{\mathbb{E}}[-y\mid \mathbf{x}]  \quad \quad \text{and} \quad \quad \overline{\mu}=\hat{\mathbb{E}}[y\mid \mathbf{x}] .
$$
When the link function $g$ is monotonically increasing, we obtain
$$
\underline{\mu}=g^{-1}(\mathbf{x}^{\top}\bm{\beta} + \underline{\beta}_0)  \quad \quad \text{and} \quad \quad \overline{\mu}=g^{-1}(\mathbf{x}^{\top}\bm{\beta} + \overline{\beta}_0).
$$
When the link function $g$ is monotonically decreasing, we obtain
$$
\underline{\mu}=g^{-1}(\mathbf{x}^{\top}\bm{\beta} + \overline{\beta}_0)  \quad \quad \text{and} \quad \quad \overline{\mu}=g^{-1}(\mathbf{x}^{\top}\bm{\beta} + \underline{\beta}_0).
$$
\end{proposition}

\begin{remark}
If $\underline{\beta}_0=\overline{\beta}_0=\beta_0$, we have
$$
\underline{\mu}=\overline{\mu}=g^{-1}(\mathbf{x}^{\top}\bm{\beta} + \beta_0).
$$
The generalized linear model with distribution uncertainty degenerates into the classical generalized linear model.
\end{remark}

\section{Methodology}\label{sec:3}

\subsection{Likelihood function}
In what follows, we restrict our attention to the canonical link function case. Although $\underline{\beta}_0$, $\overline{\beta}_0$, and $\bm{\beta}$ are the parameters of ultimate interest, the log-likelihood is first formulated as a function of $\bm{\beta}$ and $\bm{\nu}$. Given $n$ independent observations $\{(\mathbf{x}_{i},y_{i})\}_{i=1}^{n}$, the corresponding log-likelihood function is

\begin{equation}\label{eq:l}
l(\bm{\beta}, \bm{\nu},\phi;\mathbf{X},\mathbf{y}) = \sum_{i=1}^{n}\left\{\frac{y_{i}(\mathbf{x}_{i}^{\top}\bm{\beta}+\nu_{i}) - b(\mathbf{x}_{i}^{\top}\bm{\beta}+\nu_{i})}{a(\phi)} + c(y_{i}, \phi)\right\},
\end{equation}
where $\mathbf{X}$ is the $n\times p$ matrix of covariates, $\mathbf{y}=(y_1,\ldots,y_n)^{\top}$ and $\bm{\nu}=(\nu_1,\ldots,\nu_n)^{\top}$.

Taking the first derivative of \eqref{eq:l} with respect to $\bm{\beta}$ yields
$$
\frac{\partial l}{\partial \bm{\beta}} = \frac{1}{a(\phi)}\sum_{i=1}^n \left\{ y_i - b'(\mathbf{x}_{i}^{\top}\bm{\beta}+\nu_{i}) \right\}\mathbf{x}_{i}.
$$
For each $i$, we have
$$
\frac{\partial l}{\partial \nu_{i}}  = \frac{1}{a(\phi)}\left\{y_i - b'(\mathbf{x}_{i}^{\top}\bm{\beta}+\nu_{i})\right\}.
$$

\subsection{Our approach: FEMMB}

As mentioned in Section \ref{sec:GGLM}, the random intercept $\nu$ takes values in the interval $[\underline{\beta}_0,\overline{\beta}_0]$ and has distribution uncertainty. Specifically, we consider a scenario where $\{\nu_{i}\}_{i=1}^{n}$ are generated from an unknown number $k$ of distributions. The corresponding $k$-sample setup is formalized in the following assumption.

\begin{assumption}\label{assum:ksample}
The independent observations $\{(\mathbf{x}_{i},y_{i})\}_{i=1}^{n}$ are composed of $k$ samples, each sample constituting a group. For $1\leq j\leq k$, let $n_{j}$ denote the size of the $j$-th sample, where $\sum_{j=1}^{k}n_{j}=n$. In the $j$-th sample, $\nu_{i}$ follows the degenerate distribution $\delta_{c_{j}}$, where $c_{j}\in [\underline{\beta}_0,\overline{\beta}_0]$ with $\underline{\beta}_0=\min_{1\leq j\leq k} c_{j}$, $\overline{\beta}_0=\max_{1\leq j\leq k}c_{j}$, and $-\infty <\underline{\beta}_0 \leq \overline{\beta}_0<\infty$. However, the number of samples $k$ and the exact change-point locations are unknown.
\end{assumption}

\cite{li2026robust} considered the linear regression case and developed a novel two-step method named Expectation-Maximization with Moving Block (EMMB) to estimate the model parameters. Borrowing their idea, we can treat $\{\nu_{i}\}_{i=1}^{n}$ as missing data and employ an alternating maximization procedure, which is referred to as the EM algorithm, to estimate $\bm{\beta}$. With $\hat{\bm{\beta}}_{\mathrm{EM}}$ obtained via the EM algorithm, $\underline{\beta}_0$ and $\overline{\beta}_0$ can be estimated by the moving block procedure.

Let $n_{0}$ be a small window size. For simplicity, assume that $n$ is a multiple of $n_{0}$ and set $T=n/n_{0}$. Let $\mathcal{I}_j = \{ 1 + (j-1)n_0, \dots, jn_0 \}$ denote the index set of the $j$-th window. The EM algorithm for estimating $\bm{\beta}$ is then

(a) The E-step: given $\bm{\beta}^{(s)}$, find $a_{j}^{(s+1)}$ such that
$$
\sum_{i \in \mathcal{I}_j} \left\{y_i - b'(\mathbf{x}_{i}^{\top}\bm{\beta}^{(s)}+a_{j}^{(s+1)})\right\}=0,
$$
and let $\nu_{1+(j-1)n_{0}}^{(s+1)}=\cdots=\nu_{jn_{0}}^{(s+1)}=a_{j}^{(s+1)}$, for $1\leq j\leq T$.

\medskip

(b) The M-step: given $\bm{\nu}^{(s+1)}$, find $\bm{\beta}^{(s+1)}$ such that
$$
\sum_{i=1}^n \left\{ y_i - b'(\mathbf{x}_{i}^{\top}\bm{\beta}^{(s+1)}+\nu_{i}^{(s+1)}) \right\}\mathbf{x}_{i}=\mathbf{0}.
$$
Initialize $\bm{\beta}^{(0)}$, then iterate between (a) and (b) until convergence (e.g., $\|\bm{\beta}^{(s+1)} - \bm{\beta}^{(s)}\|_{\infty} < \delta$ for a prescribed tolerance $\delta$) or the maximum number of iterations is reached, to obtain the final estimate $\hat{\bm{\beta}}_{\mathrm{EM}}$.

However, for generalized linear models such as logistic regression, the EM algorithm described above suffers from the incidental parameters problem noted by \cite{neyman1948consistent}. The estimator $\hat{\bm{\beta}}_{\mathrm{EM}}$ will be inconsistent if the window size $n_{0}$ is fixed while the number of windows $T$ goes to infinity. The nonlinear link function introduces a leading-order bias of $O(1/n_{0})$ in the maximum likelihood estimate of the intercept $a_j$ within each window. As $T\to \infty$, these biases do not average out, leading to a non-vanishing asymptotic bias in the estimator of $\bm{\beta}$ and a failure of consistency.

To mitigate this problem, we add a Firth bias correction term to the log-likelihood function. Let
\begin{equation}\label{eq:tilde-l}
\tilde{l}(\bm{\beta}, \{a_j\}_{j=1}^{T}) = \sum_{j=1}^{T}
  \sum_{i \in \mathcal{I}_j}\left\{\frac{y_{i}(\mathbf{x}_{i}^{\top}\bm{\beta}+a_j) - b(\mathbf{x}_{i}^{\top}\bm{\beta}+a_j)}{a(\phi)} + c(y_{i}, \phi)\right\}+ \frac{1}{2} \sum_{j=1}^{T}\log i_{a_j} ,
\end{equation}
where
$$
i_{a_j} = \frac{1}{a(\phi)}\sum_{i\in\mathcal{I}_j} b''(\mathbf{x}_i^{\top}\bm{\beta} + a_j).
$$
Taking the first derivative of \eqref{eq:tilde-l} with respect to $\bm{\beta}$ yields
$$
\frac{\partial \tilde{l}}{\partial \bm{\beta}} = \frac{1}{a(\phi)}\sum_{j=1}^{T}
  \sum_{i \in \mathcal{I}_j} \left\{ y_i - b'(\mathbf{x}_{i}^{\top}\bm{\beta}+a_j) \right\}\mathbf{x}_{i}+\frac{1}{2}\sum_{j=1}^{T}\frac{
  \sum_{i \in \mathcal{I}_j}
  b'''(\mathbf{x}_i^{\top}\bm{\beta} + a_j) \mathbf{x}_i
}{
  \sum_{i \in \mathcal{I}_j}
  b''(\mathbf{x}_i^{\top}\bm{\beta} + a_j)
}.
$$
For $1\leq j\leq T$, taking the first derivative of \eqref{eq:tilde-l} with respect to $a_j$ yields
$$
\frac{\partial \tilde{l}}{\partial a_j}  = \frac{1}{a(\phi)}\sum_{i \in \mathcal{I}_j}\left\{y_i - b'(\mathbf{x}_{i}^{\top}\bm{\beta}+a_j)\right\}+\frac{1}{2}\frac{
  \sum_{i \in \mathcal{I}_j}
  b'''(\mathbf{x}_i^{\top}\bm{\beta} + a_j)
}{
  \sum_{i \in \mathcal{I}_j}
  b''(\mathbf{x}_i^{\top}\bm{\beta} + a_j)
}.
$$
We now propose the following improved two-step method, which we term Firth-corrected Expectation-Maximization with Moving Block (FEMMB). In the first step, a Firth-corrected EM algorithm is employed to obtain a bias-reduced estimate of $\bm{\beta}$. In the second step, a moving block procedure is applied to estimate the upper and lower bounds $\overline{\beta}_0$ and $\underline{\beta}_0$.

\medskip

\noindent\textbf{Step 1. The Firth-corrected EM algorithm for estimating $\bm{\beta}$}

Let $n_{0}$ be a small window size. For simplicity, assume $n$ is a multiple of $n_{0}$ and set $T=n/n_{0}$. Let $\mathcal{I}_j = \{ 1 + (j-1)n_0, \dots, jn_0 \}$ denote the index set of the $j$-th window. The Firth-corrected EM algorithm for estimating $\bm{\beta}$ is then

(a) The Firth-corrected E-step: given $\bm{\beta}^{(s)}$, for $1\leq j\leq T$, find $a_{j}^{(s+1)}$ such that
$$
\sum_{i \in \mathcal{I}_j} \left\{y_i - b'(\mathbf{x}_{i}^{\top}\bm{\beta}^{(s)}+a_{j}^{(s+1)})\right\}+ \frac{a(\phi)}{2}\frac{\sum_{i \in \mathcal{I}_j}b'''(\mathbf{x}_{i}^{\top}\bm{\beta}^{(s)}+a_{j}^{(s+1)})}
{\sum_{i \in \mathcal{I}_j}b''(\mathbf{x}_{i}^{\top}\bm{\beta}^{(s)}+a_{j}^{(s+1)})}=0.
$$

\medskip

(b) The Firth-corrected M-step: given $\{a_j^{(s+1)}\}_{j=1}^{T}$, find $\bm{\beta}^{(s+1)}$ such that
$$
\sum_{j=1}^{T}\sum_{i \in \mathcal{I}_j}\left\{ y_i - b'(\mathbf{x}_i^{\top}\bm{\beta}^{(s+1)} + a_j^{(s+1)}) \right\} \mathbf{x}_i
  + \frac{a(\phi)}{2}\sum_{j=1}^{T}\mathbf{W}_j(\bm{\beta}^{(s+1)}, a_j^{(s+1)})= \mathbf{0},
$$
where
$$
\mathbf{W}_j(\bm{\beta}, a_j) =
\frac{
  \sum_{i \in \mathcal{I}_j}
  b'''(\mathbf{x}_i^{\top}\bm{\beta} + a_j) \mathbf{x}_i
}{
  \sum_{i \in \mathcal{I}_j}
  b''(\mathbf{x}_i^{\top}\bm{\beta} + a_j)
}.
$$
Initialize $\bm{\beta}^{(0)}$, then iterate between (a) and (b) until convergence (e.g., $\|\bm{\beta}^{(s+1)} - \bm{\beta}^{(s)}\|_{\infty} < \delta$ for a prescribed tolerance $\delta$) or the maximum number of iterations is reached, to obtain the final estimate $\hat{\bm{\beta}}_{\mathrm{FEM}}$.

\medskip

\noindent\textbf{Step 2. The moving block procedure for estimating $\underline{\beta}_0$ and $\overline{\beta}_0$}

The data $\{(\mathbf{x}_{i},y_{i})\}_{i=1}^{n}$ are scanned sequentially into $L=n-w+1$ overlapping moving blocks of a given block length $w$ with $w > n_{0}$. For $1\leq l \leq L$, denote the data in the $l$th block by $B_{l}=\{(\mathbf{x}_{i},y_{i})\}_{i=l}^{l+w-1}$. Given $\hat{\bm{\beta}}_{\mathrm{FEM}}$ from Step 1, for each $l$, let $\hat{\alpha}_{l}$ be the solution of
$$
\sum_{i=l}^{l+w-1} \left\{y_i - b'(\mathbf{x}_{i}^{\top}\hat{\bm{\beta}}_{\mathrm{FEM}}+\alpha_{l})\right\}=0.
$$
Let
$$
\hat{\underline{\beta}}_0=\min_{1\leq l\leq L}\hat{\alpha}_{l} \qquad \text{ and }\qquad \hat{\overline{\beta}}_0=\max_{1\leq l\leq L}\hat{\alpha}_{l}.
$$

The choice of $w$ involves a bias-variance trade-off: a larger $w$ reduces the variance of each $\hat{\alpha}_{l}$ but increases the risk that a block spans multiple change points, which introduces bias; a smaller $w$ makes each block more likely to lie within a homogeneous segment but yields noisier estimates. In practice, $w$ can be selected based on domain-specific knowledge, or by data-driven methods such as cross-validation.

We could in principle apply the Firth correction to each block-specific intercept estimate. However, the first-order bias of the ordinary MLE $\hat{\alpha}_{l}$ is of order $O(1/w)$, while its standard deviation is of order $O_p(1/\sqrt{w})$. For moderate block lengths (e.g., $w=100$), the overall error of the extremum estimators is dominated by sampling variance, so additional bias correction yields only a negligible improvement. Therefore, we use ordinary MLE to estimate each block-specific intercept.

Finally, we summarize the complete FEMMB approach in Algorithm~\ref{alg:FEMMB}.

\begin{algorithm}
\caption{FEMMB approach}
\label{alg:FEMMB}
\begin{algorithmic}[1]
\Statex \textbf{Input:} data $\{(\mathbf{x}_{i},y_{i})\}_{i=1}^{n}$, small window size $n_{0}$, maximum number of iterations $max\_iter$, convergence tolerance $\delta$, block length $w$ ($w > n_0$)

\State \textbf{Initialize:} $\bm{\beta} \gets \bm{\beta}^{(0)}$
\State $T \gets  n / n_0$ ($n$ is a multiple of $n_{0}$)

\For{$iter = 1$ \textbf{to} $max\_iter$}

    \State $\bm{\beta}_{old} \gets \bm{\beta}$
    \For{$j = 1$ \textbf{to} $T$}
         \State Solve for $a_{j}$ satisfying:
         $$
          \sum_{i \in \mathcal{I}_j}\left\{y_i - b'(\mathbf{x}_{i}^{\top}\bm{\beta}+a_{j})\right\}+ \frac{a(\phi)}{2}\frac{\sum_{i \in \mathcal{I}_j}b'''(\mathbf{x}_{i}^{\top}\bm{\beta}+a_{j})}{\sum_{i \in \mathcal{I}_j}b''( \mathbf{x}_{i}^{\top}\bm{\beta}+a_{j})}=0
         $$
    \EndFor
    \State Solve for $\bm{\beta}$ satisfying:
    $$
    \sum_{j=1}^{T}\sum_{i \in \mathcal{I}_j}\left\{ y_i - b'(\mathbf{x}_{i}^{\top}\bm{\beta}+a_{j}) \right\} \mathbf{x}_i
     + \frac{a(\phi)}{2}\sum_{j=1}^{T}\frac{\sum_{i \in \mathcal{I}_j}b'''(\mathbf{x}_i^{\top}\bm{\beta} + a_j) \mathbf{x}_i}{
    \sum_{i \in \mathcal{I}_j}b''(\mathbf{x}_i^{\top}\bm{\beta} + a_j)}= \mathbf{0}
    $$

    \If{$\|\bm{\beta} - \bm{\beta}_{old}\|_{\infty} < \delta$}
        \State \textbf{break}
    \EndIf
\EndFor
\State $\hat{\bm{\beta}}_{\mathrm{FEM}} \gets \bm{\beta}$
\State $L \gets n-w+1$
\For{$l = 1$ \textbf{to} $L$}
      \State Solve for $\hat{\alpha}_{l}$ satisfying: $\displaystyle\sum_{i=l}^{l+w-1} \left\{y_i - b'(\mathbf{x}_{i}^{\top}\hat{\bm{\beta}}_{\mathrm{FEM}}+\hat{\alpha}_{l})\right\}=0$
\EndFor
\State $\hat{\overline{\beta}}_0 \gets \max_{1\leq l\leq L}\hat{\alpha}_{l}$
\State $\hat{\underline{\beta}}_0 \gets \min_{1\leq l\leq L}\hat{\alpha}_{l}$

\Statex \textbf{Output:} $\hat{\bm{\beta}}_{\mathrm{FEM}}$, $\hat{\underline{\beta}}_0$ and $\hat{\overline{\beta}}_0$
\end{algorithmic}
\end{algorithm}

\newpage

\section{Theory}\label{sec:4}

We first impose some regularity assumptions for theoretical analysis.

\begin{assumption}\label{assum:convexity}
The function $b(\cdot)$ is four times continuously differentiable, and $b''(\theta)>0$ for all $\theta$.
\end{assumption}

\begin{assumption}\label{assum:compact}
The parameter space $\mathcal{B}$ for $\bm{\beta}$ is compact and convex, and the true parameter $\bm{\beta}_0$ lies in the interior of $\mathcal{B}$.
\end{assumption}

\begin{assumption}\label{assum:n0}
For $1\leq j\leq k$, $n_{j}$ is divisible by the small window size $n_{0}$.
\end{assumption}

\begin{assumption}\label{assum:x}
The covariates $\{\mathbf{x}_i\}_{i=1}^{n}$ are independent and identically distributed (i.i.d.). The random vector $\mathbf{x}$ is sub-Gaussian, that is, for any $\bm a\in\mathbb{R}^p$, $\bm a^\top (\mathbf{x}-\mathbb{E}[\mathbf{x}])$ is a zero-mean $\kappa_u \|\bm a\|_2^2$-sub-Gaussian random variable, where $\kappa_u > 0$ is a constant. The covariance matrix $\bm{\Sigma}_{\mathbf{xx}} = \mathrm{Var}(\mathbf{x})$ is positive definite. Moreover, $\mathbf{x}$ and $\nu$ are independent.
\end{assumption}

\begin{assumption}\label{assum:infinity}
As $n \to \infty$, $n_0 \to \infty$, $T \to \infty$, and $n_{0} / T^{1/3} \to \infty$.
\end{assumption}

Assumption \ref{assum:n0} guarantees that each window of size $n_0$ lies entirely within a single underlying sample, so that all $\nu_i$ inside a given window take the same constant value. For linear and logistic regression, $b''(\cdot)$ is uniformly bounded over $\mathbb{R}$, thus the sub-Gaussian assumption on the covariates can be relaxed to $\mathbb{E}[\|\mathbf{x}\|^3] < \infty$. We maintain the sub-Gaussian assumption throughout the paper to accommodate Poisson regression, which requires additional tail control.

We now establish the convergence of the Firth-corrected EM algorithm. The following proposition shows that the algorithm converges to the unique global maximizer of the penalized likelihood under some regularity conditions.

\begin{proposition}[Convergence of the Firth-corrected EM Algorithm]\label{prop:fem_convergence}
 Let $\mathcal{A}\subset\mathbb{R}$ be a sufficiently large compact interval so that all true intercepts $\{a_{j,0}\}_{j=1}^{T}$ lie in its interior. Under Assumptions \ref{assum:ksample}--\ref{assum:x}, we further assume that the $n\times(p+T)$ design matrix is full rank, then there exists an integer $N_0$ such that for all $n_0 \ge N_0$, the Firth penalized log-likelihood $\tilde{l}(\cdot)$ is strictly concave on $\mathcal{B}\times\mathcal{A}^{T}$. For any initial value $\bm{\beta}^{(0)}\in \mathcal{B}$, the sequence $\{(\bm{\beta}^{(s)},\{a_j^{(s)}\}_{j=1}^{T})\}_{s \geq 1}$ generated by the Firth-corrected EM algorithm converges to the unique global maximum point of $\tilde{l}(\cdot)$, which can be denoted as $(\hat{\bm{\beta}}_{\mathrm{FEM}},\{\hat{a}_j\}_{j=1}^{T})$.
\end{proposition}

\begin{remark}
For logistic and Poisson regression, under the given regularity conditions, $\tilde{l}(\cdot)$ is strictly concave for any $n_0\geq\max\{2,1+p/T\}$. This result can be verified directly from the Hessian expressions in the proof of Proposition~\ref{prop:fem_convergence}. A detailed derivation is omitted for brevity.
\end{remark}

Next, we establish the large-sample properties of the Firth-corrected EM estimator $\hat{\bm{\beta}}_{\mathrm{FEM}}$, including consistency and asymptotic normality.

\begin{theorem}[Consistency of $\hat{\bm{\beta}}_{\mathrm{FEM}}$]\label{thm:consistencyFEM}
For the generalized linear model
$$
f(y; \theta, \phi) = \exp\Bigl(\frac{y\theta - b(\theta)}{a(\phi)} + c(y, \phi)\Bigr),
$$
where $\theta=\eta = \nu + \mathbf{x}^{\top}\bm{\beta}$. Under Assumptions \ref{assum:ksample}--\ref{assum:infinity}, the Firth-corrected EM
estimator $\hat{\bm{\beta}}_{\mathrm{FEM}}$ is consistent for the true parameter $\bm{\beta}_0$, i.e.,
$$
\hat{\bm{\beta}}_{\mathrm{FEM}} \xrightarrow{P} \bm{\beta}_0 \qquad \text{ as } n \to \infty.
$$
\end{theorem}

\begin{remark}
For consistency, Assumption \ref{assum:infinity} can be relaxed. It suffices that $n_0 \to \infty$ and $T \to \infty$ as $n \to \infty$. The stronger rate condition $n_{0} / T^{1/3} \to \infty$ is only needed for asymptotic normality.
\end{remark}

\begin{theorem}[Asymptotic Normality of $\hat{\bm{\beta}}_{\mathrm{FEM}}$]\label{thm:asymnormalFEM}
For the generalized linear model
$$
f(y; \theta, \phi) = \exp\Bigl(\frac{y\theta - b(\theta)}{a(\phi)} + c(y, \phi)\Bigr),
$$
where $\theta=\eta = \nu + \mathbf{x}^{\top}\bm{\beta}$. Under Assumptions \ref{assum:ksample}--\ref{assum:infinity}, the Firth-corrected EM estimator $\hat{\bm{\beta}}_{\mathrm{FEM}}$ satisfies
$$
\sqrt{n}\bigl(\hat{\bm{\beta}}_{\mathrm{FEM}} - \bm{\beta}_0\bigr)\xrightarrow{d} \mathcal{N}(\mathbf{0}, a(\phi)\mathbf{J}^{-1}).
$$
where
$$
\mathbf{J} = \lim_{T\to\infty} \frac{1}{T}\sum_{j=1}^T \Bigl(
\mathbb{E}\bigl[ b_{ij}'' \mathbf{x}_i \mathbf{x}_i^\top \bigr]
- \frac{ \mathbb{E}[ b_{ij}'' \mathbf{x}_i ] \mathbb{E}[ b_{ij}'' \mathbf{x}_i ]^\top }
       { \mathbb{E}[ b_{ij}'' ] }
\Bigr),
$$
with $b_{ij}'' = b''(a_{j,0} + \mathbf{x}_i^\top\bm{\beta}_0)$.
\end{theorem}

To establish the consistency of the FEMMB estimators $\hat{\overline{\beta}}_0$ and $\hat{\underline{\beta}}_0$, we impose the following regularity condition on the block length $w$.

\begin{assumption}\label{assum:w}
The block length $w$ satisfies $w > n_0$ and $w \le \min_{1 \le j \le k} n_j$. Furthermore, as $n \to \infty$, $w \to \infty$ and $w / \log n \to \infty$.
\end{assumption}

\begin{theorem}[Consistency of $\hat{\overline{\beta}}_0$ and $\hat{\underline{\beta}}_0$]\label{thm:consistency_bounds}
Under Assumptions \ref{assum:ksample}--\ref{assum:w}, the FEMMB estimators $\hat{\overline{\beta}}_0$ and $\hat{\underline{\beta}}_0$ are consistent for the upper bound $\overline{\beta}_0$ and lower bound $\underline{\beta}_0$, respectively, i.e.,
$$
\hat{\overline{\beta}}_0 \xrightarrow{P} \overline{\beta}_0 \quad \text{and} \quad \hat{\underline{\beta}}_0 \xrightarrow{P} \underline{\beta}_0\quad \text{ as } n \to \infty.
$$
\end{theorem}

\begin{remark}
Let $n^{*}$ denote the maximum size among the homogeneous samples with intercept $\overline{\beta}_0$, and $n_{*}$ denote the maximum size among the homogeneous samples with intercept $\underline{\beta}_0$. The condition $w \le \min_{1 \le j \le k} n_j$ can be relaxed to $w \leq \min\{n^{*},n_{*}\}$.
\end{remark}

\section{Simulation studies}\label{sec:5}

In this section, we investigate the finite-sample performance of the proposed FEMMB method and compare it with the EMMB method and pooled maximum likelihood estimation (pooled MLE), which ignores intercept heterogeneity and treats all intercepts as identical. We use bias and mean squared error (MSE) of the regression coefficient estimators as evaluation metrics. In addition, we report the means of the lower and upper intercept estimators $\hat{\underline{\beta}}_0$ and $\hat{\overline{\beta}}_0$ obtained from FEMMB and EMMB, as well as the mean of the single constant intercept estimator $\hat{\beta}_{0}$ obtained from pooled MLE.

\begin{example}[Logistic Regression]\label{exam:1}
In this example, we simulated 500 datasets consisting of $n$ observations from the model $y\sim\text{Bernoulli}\{p(\nu + \mathbf{x}^{\top}\bm{\beta})\}$, where $p(u)=1/(1+\exp(-u))$, $\bm{\beta}=(1.5,-0.6)^{\top}$, and the components of $\mathbf{x}$ are independent and identically distributed as $N(0,1)$. Here, $\nu$ is a random intercept and takes values in the interval $[\underline{\beta}_0,\overline{\beta}_0]$, where $\underline{\beta}_0=-3$ and $\overline{\beta}_0=1$. Specifically, $\{\nu_{i}\}_{i=1}^{n}$ are generated as follows. First, let $c_{1}=-3, c_{k}=1$ and $c_{j}$ follow the uniform distribution $U(-3,1)$ for $2\leq j\leq k-1$. Second, let $\nu_{i}$ follow the degenerate distribution $\delta_{c_{j}}$ for $1\leq j\leq k, 1+(j-1)m\leq i\leq jm$ and thus $n=km$. The numerical
results associated with this example are reported in Tables \ref{table:1} and \ref{table:2}.
\end{example}

\begin{table}[tbp]
\centering
\caption{Simulation results for logistic regression with $k=2$ and varying $m$}
\label{table:1}
\begin{tabular}{l cc cc c}
\toprule
\multirow{2}{*}{Method} & \multicolumn{2}{c}{$\hat{\beta}_{1}$} & \multicolumn{2}{c}{$\hat{\beta}_{2}$} & $(\hat{\underline{\beta_0}},\hat{\overline{\beta_0}})$ \\
\cmidrule(lr){2-3} \cmidrule(lr){4-5} \cmidrule(lr){6-6}
 & Bias & MSE & Bias & MSE & Mean \\
\midrule
\multicolumn{6}{l}{$(n,w,n_{0})=(400,180,20)$} \\
FEMMB & $ 0.0187$ & $0.0387$  & $-0.0014$ & $0.0270$  & $(-3.1241, 1.0757)$  \\
EMMB & $0.1419$ & $0.0682$  & $-0.0502$ & $0.0347$  & $(-3.2480, 1.1208)$  \\
Pooled MLE & $-0.6423$ & $0.4290$  & $0.2657$ & $0.0845$ & $-0.5532$  \\[0.3em]
\multicolumn{6}{l}{$(n,w,n_{0})=(1000,480,20)$} \\
FEMMB & $0.0011$ & $0.0173$  &  $0.0064$ & $0.0098$  & $(-3.0450, 1.0216)$  \\
EMMB & $0.1215$ & $0.0364$  &  $-0.0412$ & $0.0133$  & $(-3.1633, 1.0636)$  \\
Pooled MLE & $-0.6537$ & $0.4335$  &  $0.2615$ &  $0.0733$ & $-0.5559$  \\[0.3em]
\multicolumn{6}{l}{$(n,w,n_{0})=(2000,980,20)$} \\
FEMMB & $ 0.0074$ & $0.0098$  & $-0.0005$ & $0.0055$  & $(-3.0295, 1.0172)$ \\
EMMB & $0.1279$ & $0.0285$  & $-0.0485$ & $0.0090$  & $(-3.1471, 1.0591)$ \\
Pooled MLE & $-0.6542$ & $0.4317$  & $0.2645$ & $0.0724$ & -0.5536\\
\bottomrule
\end{tabular}
\end{table}

\begin{table}[tbp]
\centering
\caption{Simulation results for logistic regression with $m=200$ and varying $k$}
\label{table:2}
\begin{tabular}{l cc cc c}
\toprule
\multirow{2}{*}{Method} & \multicolumn{2}{c}{$\hat{\beta}_{1}$} & \multicolumn{2}{c}{$\hat{\beta}_{2}$} & $(\hat{\underline{\beta_0}},\hat{\overline{\beta_0}})$ \\
\cmidrule(lr){2-3} \cmidrule(lr){4-5} \cmidrule(lr){6-6}
 & Bias & MSE & Bias & MSE & Mean \\
\midrule
\multicolumn{6}{l}{$(n,w,n_{0})=(400,180,20)$} \\
FEMMB & $ 0.0187$ & $0.0387$  & $-0.0014$ & $0.0270$  & $(-3.1241, 1.0757)$  \\
EMMB & $0.1419$ & $0.0682$  & $-0.0502$ & $0.0347$  & $(-3.2480, 1.1208)$  \\
Pooled MLE & $-0.6423$ & $0.4290$  & $0.2657$ & $0.0845$ & $-0.5532$  \\[0.3em]
\multicolumn{6}{l}{$(n,w,n_{0})=(1000,180,20)$} \\
FEMMB & $0.0038$ & $0.0141$  &  $ -0.0003$ & $0.0106$  & $(-3.1681, 1.0916)$  \\
EMMB & $0.1157$ & $0.0306$  &  $-0.0449$ & $0.0144$  & $(-3.2811, 1.1332)$  \\
Pooled MLE & $-0.4330$  & $0.1995$   &  $0.1760$ &  $0.0379$ & $-0.7243$  \\[0.3em]
\multicolumn{6}{l}{$(n,w,n_{0})=(2000,180,20)$} \\
FEMMB & $ 0.0033$ & $0.0072$  &  $-0.0006$ & $0.0042$  & $(-3.1961, 1.0977)$ \\
EMMB & $0.1117$ & $0.0212$  &  $-0.0438$ & $0.0067$  & $(-3.3054, 1.1378)$ \\
Pooled MLE & $-0.3524$ & $0.1313$  & $0.1407$ & $0.0233$ & $-0.7560$\\
\bottomrule
\end{tabular}
\end{table}

As shown in Table \ref{table:1}, for the setting with $k=2$ and varying $m$, the proposed FEMMB method substantially outperforms the EMMB method and pooled MLE. As $m$ increases, both the bias and MSE of FEMMB decrease further. The EMMB method shows some improvement in MSE but its bias remains large, reflecting the severe incidental parameter problem in logistic regression. The pooled MLE, which completely ignores intercept heterogeneity, exhibits virtually no improvement and suffers persistent large estimation bias. From Table \ref{table:2}, we can see that fixing $m=200$ and increasing $k$ yields performance gains for all three methods. Nevertheless, FEMMB still achieves the smallest bias and MSE among the three methods.

\begin{example}[Poisson Regression]\label{exam:2}
In this example, we simulated 500 datasets consisting of $n$ observations from the model $y\sim\text{Poisson}\{\exp(\nu + \mathbf{x}^{\top}\bm{\beta})\}$, where $\bm{\beta}, \mathbf{x}$ and $\nu$ are the same as those in Example \ref{exam:1}. The numerical
results associated with this example are reported in Tables \ref{table:3} and \ref{table:4}.
\end{example}

\begin{table}[tbp]
\centering
\caption{Simulation results for Poisson regression with $k=2$ and varying $m$}
\label{table:3}
\begin{tabular}{l cc cc c}
\toprule
\multirow{2}{*}{Method} & \multicolumn{2}{c}{$\hat{\beta}_{1}$} & \multicolumn{2}{c}{$\hat{\beta}_{2}$} & $(\hat{\underline{\beta_0}},\hat{\overline{\beta_0}})$ \\
\cmidrule(lr){2-3} \cmidrule(lr){4-5} \cmidrule(lr){6-6}
 & Bias & MSE & Bias & MSE & Mean \\
\midrule
\multicolumn{6}{l}{$(n,w,n_{0})=(400,180,20)$} \\
FEMMB & $-0.0002$ & $0.0010$  & $0.0003$ & $0.0008$  & $(-3.0705, 1.0070)$   \\
EMMB &  $-0.0002$ & $0.0010$  & $0.0003$ & $0.0008$  & $(-3.0705, 1.0070)$  \\
Pooled MLE & $-0.0227$ & $0.0351$  & $0.0080$ & $0.0206$ & $0.3227$  \\[0.3em]
\multicolumn{6}{l}{$(n,w,n_{0})=(1000,480,20)$} \\
FEMMB & $0.0018$ & $0.0004$  &  $0.0003$ & $0.0003$  & $(-3.0301, 1.0008)$  \\
EMMB &  $0.0018$ & $0.0004$  &  $0.0003$ & $0.0003$  & $(-3.0301, 1.0008)$  \\
Pooled MLE & $-0.0060$ & $0.0170$  &  $0.0059$ &  $0.0092$ & $0.3240$  \\[0.3em]
\multicolumn{6}{l}{$(n,w,n_{0})=(2000,980,20)$} \\
FEMMB & $0.0007$ & $0.0002$  & $0$ & $0.0001$  & $(-3.0165, 1.0004)$ \\
EMMB & $0.0007$ & $0.0002$  & $0$ & $0.0001$  & $(-3.0165, 1.0004)$ \\
Pooled MLE & $-0.0091$ & $0.0102$  & $0.0020$ & $0.0046$ & 0.3308\\
\bottomrule
\end{tabular}
\end{table}

\begin{table}[tbp]
\centering
\caption{Simulation results for Poisson regression with $m=200$ and varying $k$}
\label{table:4}
\begin{tabular}{l cc cc c}
\toprule
\multirow{2}{*}{Method} & \multicolumn{2}{c}{$\hat{\beta}_{1}$} & \multicolumn{2}{c}{$\hat{\beta}_{2}$} & $(\hat{\underline{\beta_0}},\hat{\overline{\beta_0}})$ \\
\cmidrule(lr){2-3} \cmidrule(lr){4-5} \cmidrule(lr){6-6}
 & Bias & MSE & Bias & MSE & Mean \\
\midrule
\multicolumn{6}{l}{$(n,w,n_{0})=(400,180,20)$} \\
FEMMB & $-0.0002$ & $0.0010$  & $0.0003$ & $0.0008$  & $(-3.0705, 1.0070)$  \\
EMMB & $-0.0002$ & $0.0010$  & $0.0003$ & $0.0008$  & $(-3.0705, 1.0070)$  \\
Pooled MLE & $-0.0227$ & $0.0351$  & $0.0080$ & $0.0206$ & $0.3227$  \\[0.3em]
\multicolumn{6}{l}{$(n,w,n_{0})=(1000,180,20)$} \\
FEMMB & $-0.0001$ & $0.0006$  &  $0$ & $0.0005$  & $(-3.0906, 1.005)$  \\
EMMB & $-0.0001$ & $0.0006$  &  $0$ & $0.0005$  & $(-3.0906, 1.005)$  \\
Pooled MLE & $-0.0043$  & $0.0230$   &  $0.0109$ &  $0.0147$ & $-0.1048$  \\[0.3em]
\multicolumn{6}{l}{$(n,w,n_{0})=(2000,180,20)$} \\
FEMMB & $ 0.0008$ & $0.0004$  &  $-0.0004$ & $0.0003$  & $(-3.1210, 1.0062)$  \\
EMMB & $ 0.0008$ & $0.0004$  &  $-0.0004$ & $0.0003$  & $(-3.1210, 1.0062)$ \\
Pooled MLE & $-0.0113$ & $0.0139$  & $0.0024$ & $0.0071$ & $-0.2485$\\
\bottomrule
\end{tabular}
\end{table}

From Table \ref{table:3}, we can see that for the setting of $k=2$ and varying $m$, FEMMB and EMMB have exactly the same performance. Both exhibit virtually no bias and very small MSE, while pooled MLE performs worst. As the total sample size $n$ increases, the performance of all three methods improves. Table \ref{table:4}, corresponding to $m=200$ and varying $k$, reveals similar patterns. It is noteworthy that for Poisson regression, EMMB does not suffer from the incidental-parameter problem: the structure of the Poisson score equations ensures that the biases of the estimated window-specific intercepts cancel out in the M-step, leading to a consistent estimator for the regression coefficients. This explains why FEMMB and EMMB coincide.

The two simulation examples illustrate the effectiveness of the proposed FEMMB method. For logistic regression,  where EMMB suffers from the severe incidental-parameter problem, FEMMB substantially reduces the estimation bias and MSE. For Poisson regression, where the EMMB estimator is already consistent, FEMMB coincides with EMMB. Both methods substantially outperform pooled MLE. Furthermore, both EMMB and FEMMB recover the true intercept range by the moving block method, and the accuracy of this range improves as $w$ grows, while pooled MLE provides only a single point estimate. These results demonstrate that modeling the intercept heterogeneity through a piecewise constant structure is inherently valuable, and that the Firth correction is effective when needed while incurring no additional costs otherwise.

We also find that FEMMB exhibits greater numerical stability than EMMB in the estimation of window-specific intercepts. For small $n_{0}$, standard root-finding algorithms such as Brent's method or Bisection method may fail or produce unreliable solutions for the E-step of EMMB, whereas the Firth correction ensures a well-behaved estimating equation for the E-step of FEMMB that is easily solved by standard methods. To ensure a fair comparison, we adopted Newton's method for the E-step of both EMMB and FEMMB in our simulations. Newton's method can provide stable solutions for EMMB even for small $n_{0}$, although it is unnecessary for FEMMB.

\section{Real data analysis}\label{sec:6}

In this section, we investigate the Default of Credit Card Clients dataset from \cite{Yeh2009TheCO}. The binary response variable is default payment, where $Y=1$ denotes that a cardholder defaulted on their payment in October 2005, and $Y=0$ otherwise. The dataset contains 30,000 observations, of which 6,636 (22.12\%) correspond to defaulted clients, showing moderate class imbalance. There are 23 explanatory variables in the dataset, including amount of the given credit, demographic features (gender, education, marital status, age), and six consecutive months of financial records from April to September 2005: repayment status, amount of bill statement, and amount of payment.

We use 1/3 of the data for training and the remaining 2/3 for testing, with stratification based on the response variable. We apply one-hot encoding to the three categorical variables (gender, education, and marital status) and drop the first category to avoid the dummy variable trap. Categories are specified according to the original data dictionary. All remaining numerical explanatory variables are standardized. All preprocessing transformations are fit exclusively on the training set and then applied to both training and test sets. We assume that cardholders with the same or similar amounts of the given credit share the same intercept. To align the dataset with our model specification (Assumption \ref{assum:ksample}), we sort the training set by this variable in ascending order and dropped it from both the training and test sets.

We evaluate model performance using four metrics: precision, recall, F1-score, and balanced accuracy. For this credit card default prediction task, the positive class corresponds to default customers (i.e., samples labeled 1), and all reported precision, recall, and F1-score are computed for this positive class. For comparison, we benchmark our FEMMB method against several competing approaches: the NP umbrella algorithm of \cite{tong2018neyman}, SMOTE \citep{chawla2002smote}, random under-sampling (RUS), and pooled logistic regression (pooled LR). To ensure a fair comparison, logistic regression is adopted as the base classifier for all competing methods. Guided  by Assumptions \ref{assum:infinity} and \ref{assum:w}, we set $n_0=50, w=100$ for the FEMMB method. For the NP umbrella algorithm, we set $\alpha=0.1, \delta=0.1$.

In our FEMMB method, we obtain $\hat{\bm{\beta}}_{\mathrm{FEM}}$, $\hat{\underline{\beta}}_0$, and $\hat{\overline{\beta}}_0$ during training. For prediction, we use $\mathbf{x}^{\top}\hat{\bm{\beta}}_{\mathrm{FEM}}+\hat{\overline{\beta}}_0$ as the linear predictor. The conservative choice of the upper intercept aligns the method with the practical requirement of cautious risk assessment in credit default prediction. Since the NP umbrella algorithm places the default class under the null hypothesis, we invert the training labels for this method (i.e., $y_{train, flip}=1- y_{train}$) and map the test predictions back to the original label encoding before computing evaluation metrics against the untouched $y_{test}$. The numerical results are summarized in Table \ref{table:real}.

\begin{table}[tbp]
\centering
\caption{Classification performance on Default of Credit Card Clients data}
\label{table:real}
\begin{tabular}{l c c c c}
\toprule
 Method & Precision & Recall & F1-score & Balanced Accuracy\\
\midrule
FEMMB &0.5781 & $0.4702$  & $\textbf{0.5186}$ & $\textbf{0.6864}$  \\
NP(LR-based) & $0.2477$ & $ \textbf{0.9146}$  & $0.3898$ & $0.5628$ \\
SMOTE $+$ LR & $0.4196$ & $0.6103$  & $0.4973$ & $0.6853$ \\
RUS $+$ LR& $0.4231$ & $0.6004$  & $0.4964$ & $0.6839$ \\
Pooled LR & $\textbf{0.6913}$ & $0.2724$  & $0.3908$ & $0.6189$ \\
\bottomrule
\end{tabular}
\end{table}

From Table \ref{table:real}, we can see that pooled logistic regression achieves the highest precision of 0.6913 but suffers a very low recall of 0.2724, indicating that it misses the majority of true defaulters. Conversely, the NP umbrella algorithm attains the highest recall of 0.9146 at the expense of poor precision (0.2477), producing excessive false positive alerts that misclassify numerous normal clients as high-risk defaulters. Resampling approaches SMOTE and RUS yield moderate improvements in the trade-off between precision and recall by rebalancing the training data. Our proposed FEMMB method outperforms all competing methods in both F1-score (0.5186) and balanced accuracy (0.6864), achieving a more favorable balance between identifying real defaulters and controlling misclassification of normal clients.

\section{Conclusion}\label{sec:7}

Inspired by the theory of sublinear expectation, we investigated generalized linear models under distribution uncertainty, where the random intercept $\nu$ has mean uncertainty and takes values in an unknown bounded interval $[\underline{\beta}_0,\overline{\beta}_0]$. We developed a novel two-step method, named Firth-corrected Expectation-Maximization with Moving Block (FEMMB), to estimate the unknown parameters. The Firth correction effectively mitigates the incidental parameters problem that arises in GLMs with many window-specific intercepts. Theoretical properties of the obtained estimators are established under mild regularity conditions. We demonstrated the effectiveness of the proposed FEMMB method via simulation studies and a real-data example.

There are several promising directions for future research. The first interesting problem is to construct hypothesis testing procedures for the resulting estimators. Another direction is to consider more complex models that also accommodate distribution uncertainty in the regression coefficients.

\appendix

\section{Proofs}

\begin{proof}[Proof of Proposition \ref{prop:glm-interval}]
When $g$ is monotonically increasing, we have
\begin{equation}\nonumber
\begin{split}
\overline{\mu}&=\hat{\mathbb{E}}[y\mid \mathbf{x}] \\
 &=\sup_{\theta \in \Theta} \mathbb{E}_{F_{\theta}}[y\mid \mathbf{x}]\\
 &=\sup_{\theta \in \Theta} \mathbb{E}_{F_{\theta}}\left[\mathbb{E}[y\mid \mathbf{x},\nu]\mid \mathbf{x}\right]\\
 &=\sup_{\theta \in \Theta} \mathbb{E}_{F_{\theta}}[g^{-1}(\mathbf{x}^{\top}\bm{\beta} + \nu)\mid \mathbf{x}]\\
 &= \sup_{\theta\in\Theta} \int_{\underline{\beta}_0}^{\overline{\beta}_0} g^{-1}(\mathbf{x}^\top \bm{\beta} + u) dF_{\theta}(u)\\
 &=g^{-1}(\mathbf{x}^{\top}\bm{\beta} + \overline{\beta}_0).
\end{split}
\end{equation}
Here, with a slight abuse of notation, $F_{\theta}$ denotes a possible distribution of the random variable (vector) inside the expectation operator. The third equality follows from the law of iterated expectations. The last equality holds because $ g^{-1}$ is monotonically increasing and the degenerate distribution $\delta_{\overline{\beta}_0}$ is a possible distribution of $\nu$. Similarly,
\begin{equation}\nonumber
\begin{split}
\underline{\mu}&=-\hat{\mathbb{E}}[-y\mid \mathbf{x}] \\
 &=\inf_{\theta \in \Theta} \mathbb{E}_{F_{\theta}}[y\mid \mathbf{x}]\\
 &=g^{-1}(\mathbf{x}^{\top}\bm{\beta} + \underline{\beta}_0).
\end{split}
\end{equation}
For the case where $g$ is monotonically decreasing, analogous arguments yield
$$
\underline{\mu}=g^{-1}(\mathbf{x}^{\top}\bm{\beta} + \overline{\beta}_0)   \quad \quad \text{and} \quad \quad \overline{\mu}=g^{-1}(\mathbf{x}^{\top}\bm{\beta} + \underline{\beta}_0).
$$
This completes the proof.
\end{proof}

\medskip

\begin{proof}[Proof of Proposition \ref{prop:fem_convergence}]
Since $b''(\cdot) > 0$, the design matrix is full rank, and the covariance matrix $\bm{\Sigma}_{\mathbf{xx}}$ is positive definite, the original log-likelihood function
$$
l(\bm{\beta}, \{a_j\}_{j=1}^{T}) =\sum_{j=1}^{T}
  \sum_{i \in \mathcal{I}_j}\left\{\frac{y_{i}(\mathbf{x}_{i}^{\top}\bm{\beta}+a_j) - b(\mathbf{x}_{i}^{\top}\bm{\beta}+a_j)}{a(\phi)} + c(y_{i}, \phi)\right\}.
$$
is strictly concave on $\mathcal{B}\times\mathcal{A}^{T}$.

The Firth penalized log-likelihood function is
$$
\tilde{l}(\bm{\beta}, \{a_j\}_{j=1}^{T}) = l(\bm{\beta}, \{a_j\}_{j=1}^{T}) + \frac{1}{2} \sum_{j=1}^{T}\log i_{a_j} ,
$$
where
$$
i_{a_j} = \frac{1}{a(\phi)}\sum_{i\in\mathcal{I}_j} b''(\mathbf{x}_i^{\top}\bm{\beta} + a_j).
$$
The second-order derivatives of $\tilde{l}$ are given by
$$
\frac{\partial^{2}\tilde{l}}{\partial \bm{\beta} \partial \bm{\beta}^{\top}}=-\sum_{j=1}^T\Bigl\{\frac{1}{a(\phi)}\sum_{i\in\mathcal{I}_j} b''(\mathbf{x}_{i}^{\top}\bm{\beta}+a_j)\mathbf{x}_i\mathbf{x}_i^\top-\frac{1}{2}\frac{(\sum b''''\mathbf{x}_i\mathbf{x}_i^\top)(\sum b'')
-(\sum b''' \mathbf{x}_i)(\sum b''' \mathbf{x}_i)^{\top}}
{(\sum b'')^{2}}\Bigr\},
$$

$$
\frac{\partial^{2}\tilde{l}}{\partial \bm{\beta} \partial a_j}=-\frac{1}{a(\phi)}\sum_{i\in\mathcal{I}_j} b''(\mathbf{x}_{i}^{\top}\bm{\beta}+a_j)\mathbf{x}_i+\frac{1}{2}\frac{(\sum b''''\mathbf{x}_i)(\sum b'')
-(\sum b''' \mathbf{x}_i)(\sum b''')}
{(\sum b'')^{2}},
$$

$$
\frac{\partial^{2}\tilde{l}}{\partial a_{j}^{2}}=-\frac{1}{a(\phi)}\sum_{i\in\mathcal{I}_j} b''(\mathbf{x}_{i}^{\top}\bm{\beta}+a_j)+\frac{1}{2}\frac{(\sum b'''')(\sum b'')
-(\sum b''' )^{2}}
{(\sum b'')^{2}}.
$$
Since the Firth correction term induces a perturbation of magnitude $O(1/n_0)$ relative to the Hessian of the original log-likelihood, there exists an integer $N_0$ such that for all $n_0 \ge N_0$, the penalized log-likelihood $\tilde{l}(\bm{\beta}, \{a_j\}_{j=1}^{T})$ is strictly concave on $\mathcal{B}\times\mathcal{A}^{T}$ and has a unique global maximizer.

The Firth-corrected EM algorithm proceeds by block coordinate ascent: The Firth-corrected E-step maximizes $\tilde{l}$ with respect to $\{a_j\}_{j=1}^{T}$ for fixed $\bm{\beta}$, and the Firth-corrected M-step maximizes $\tilde{l}$ with respect to $\bm{\beta}$ for fixed $\{a_j\}_{j=1}^{T}$. Thus,
$$
\tilde{l}(\bm{\beta}^{(s+1)}, \{a_j^{(s+1)}\}_{j=1}^{T}) \geq \tilde{l}(\bm{\beta}^{(s)}, \{a_j^{(s)}\}_{j=1}^{T}), \quad \forall s \geq 1.
$$
Since the parameter space $\mathcal{B}\times\mathcal{A}^{T}$ is compact and $\tilde{l}$ is continuous, the sequence of penalized log-likelihood values $\{\tilde{l}(\bm{\beta}^{(s)}, \{a_j^{(s)}\}_{j=1}^{T})\}_{s \geq 1}$ is bounded above and thus converges to a limit $\tilde{l}^{*}$. Strict concavity of $\tilde{l}(\cdot)$ and the exact coordinate-wise maximization property of the algorithm imply that the iterative sequence $\{(\bm{\beta}^{(s)},\{a_j^{(s)}\}_{j=1}^{T})\}_{s \geq 1}$  converges to the unique global maximum point of $\tilde{l}(\cdot)$, which can be denoted as $(\hat{\bm{\beta}}_{\mathrm{FEM}},\{\hat{a}_j\}_{j=1}^{T})$.

This completes the proof.
\end{proof}

\medskip

\begin{proof}[Proof of Theorem \ref{thm:consistencyFEM}]
For any $\bm{\beta} \in \mathcal{B}$, $n_0 \ge N_0$, and $1\leq j\leq T$, let $\hat{a}_j(\bm{\beta})$ denote the unique solution of the following equation

$$
F_{j}(a_j, \bm{\beta})=\frac{1}{n_0}\sum_{i \in \mathcal{I}_j}\left\{y_i - b'(\mathbf{x}_{i}^{\top}\bm{\beta}+a_j)\right\}+\frac{a(\phi)}{2n_0}\frac{
  \sum_{i \in \mathcal{I}_j}
  b'''(\mathbf{x}_i^{\top}\bm{\beta} + a_j)
}{
  \sum_{i \in \mathcal{I}_j}
  b''(\mathbf{x}_i^{\top}\bm{\beta} + a_j)
}=0,
$$
and let $a_j^*(\bm{\beta})$ be the unique population solution  of
$$
\mathbb{E}\left[ y_i - b'(\mathbf{x}_i^{\top}\bm{\beta}+a_j^*(\bm{\beta})) \mid j \right] = 0, \qquad i \in \mathcal{I}_j.
$$
The Firth correction term in $F_{j}(a_j, \bm{\beta})$ is introduced to eliminate the leading $O(1/n_0)$ finite-sample bias of $\hat{a}_j(\bm{\beta})$. As $n_0 \to \infty$, this correction term of order $O(1/n_0)$ vanishes asymptotically. Therefore,
$$
\hat{a}_j(\bm{\beta})\xrightarrow{P} a_j^*(\bm{\beta}) \qquad \text{ as } n_0 \to \infty.
$$
For each window $j$, the implicit function theorem implies that $\hat{a}_j(\bm{\beta})$ is continuously differentiable in $\bm{\beta}$. Moreover, its derivative is bounded in probability uniformly over the compact set $\mathcal{B}$, since $b''(\cdot)$ is bounded away from zero on compact set and $\mathbf{x}$ is sub-Gaussian. Hence $\hat{a}_j(\bm{\beta})$ is stochastically equicontinuous. Together with pointwise convergence, this yields uniform convergence in probability \citep[Lemma 2.8]{newey1994large}
$$
\sup_{\bm{\beta} \in \mathcal{B}}\left| \hat{a}_j(\bm{\beta}) - a_j^*(\bm{\beta}) \right| = o_p(1).
$$
By the weak law of large numbers applied to independent windows, we have
$$
\frac{1}{T} \sum_{j=1}^{T}\sup_{\bm{\beta} \in \mathcal{B}}\left| \hat{a}_j(\bm{\beta}) - a_j^*(\bm{\beta}) \right| = o_p(1).
$$
Define the sample estimating function
$$
\mathbf{H}_n(\bm{\beta}) = \frac{1}{n_0T} \sum_{j=1}^{T}
  \sum_{i \in \mathcal{I}_j} \left\{ y_i - b'(\mathbf{x}_{i}^{\top}\bm{\beta}+\hat{a}_j(\bm{\beta})) \right\}\mathbf{x}_{i}+\frac{a(\phi)}{2n_0T}\sum_{j=1}^{T} \frac{
  \sum_{i \in \mathcal{I}_j}
  b'''(\mathbf{x}_i^{\top}\bm{\beta} + \hat{a}_j(\bm{\beta})) \mathbf{x}_i
}{
  \sum_{i \in \mathcal{I}_j}
  b''(\mathbf{x}_i^{\top}\bm{\beta} + \hat{a}_j(\bm{\beta}))
}.
$$
By Proposition \ref{prop:fem_convergence}, $\hat{\bm{\beta}}_{\mathrm{FEM}}$ satisfies $\mathbf{H}_n(\hat{\bm{\beta}}_{\mathrm{FEM}}) = \mathbf{0}$.

For each window $j$, define the window-specific population estimating function
$$
\mathbf{m}_j(\bm{\beta})
= \mathbb{E}\left[ \left( y_i - b'(\mathbf{x}_i^{\top}\bm{\beta}+a_j^*(\bm{\beta})) \right) \mathbf{x}_i \mid j \right],
\quad i \in \mathcal{I}_j,
$$
where the expectation is taken with respect to the true distribution of observations within window $j$.

By the law of large numbers applied to independent windows, the cross-window average converges in probability as $T \to \infty$. We then define the overall population estimating function as
$$
\mathbf{M}(\bm{\beta})
= \lim_{T\to\infty} \frac{1}{T}\sum_{j=1}^T \mathbf{m}_j(\bm{\beta}),
$$
where this limit exists uniformly over $\bm{\beta} \in \mathcal{B}$ under Assumption \ref{assum:infinity}.

Let
$$
\mathbf{H}_n^{*}(\bm{\beta}) = \frac{1}{n_0T} \sum_{j=1}^{T}
  \sum_{i \in \mathcal{I}_j} \left\{ y_i - b'(\mathbf{x}_{i}^{\top}\bm{\beta}+a_j^*(\bm{\beta})) \right\}\mathbf{x}_{i}.
$$
Note that the Firth correction term in $\mathbf{H}_n(\bm{\beta})$ is of order $O(1/n_0)$ and vanishes as $n_0\to\infty$. By the mean value theorem applied to $b'(\cdot)$, we have
$$
\sup_{\bm{\beta} \in \mathcal{B}} \left\| \mathbf{H}_n(\bm{\beta}) - \mathbf{H}_n^{*}(\bm{\beta}) \right\| \leq \frac{C}{T} \sum_{j=1}^{T}\sup_{\bm{\beta} \in \mathcal{B}}\left| \hat{a}_j(\bm{\beta}) - a_j^*(\bm{\beta}) \right|+o_p(1)=o_p(1).
$$
By the uniform law of large numbers over the compact set $\mathcal{B}$, we have
$$
\sup_{\bm{\beta} \in \mathcal{B}} \left\| \mathbf{H}_n^{*}(\bm{\beta}) - \mathbf{M}(\bm{\beta}) \right\| = o_p(1).
$$
Thus
$$
\sup_{\bm{\beta} \in \mathcal{B}} \left\| \mathbf{H}_n(\bm{\beta}) - \mathbf{M}(\bm{\beta}) \right\| = o_p(1).
$$
The model is a correctly specified generalized linear model with the canonical link function. At the true structural parameter $\bm{\beta}_0$, we have $a_j^*(\bm{\beta}_0) = a_{j,0}$ and
$$
\mathbb{E}\left[ y_i - b'(\mathbf{x}_i^{\top}\bm{\beta}_0+a_{j,0}) \mid \mathbf{x}_i, j \right] = 0, \qquad i \in \mathcal{I}_j.
$$
It follows that $\mathbf{m}_j(\bm{\beta}_0) = \mathbf{0}$ for $1\leq j\leq T$, and hence $\mathbf{M}(\bm{\beta}_0) = \mathbf{0}$. Since the model is correctly specified and the covariate covariance matrix $\bm{\Sigma}_{\mathbf{xx}}$ is positive definite (Assumption~\ref{assum:x}), the population estimating equation $\mathbf{M}(\bm{\beta}) = \mathbf{0}$ has unique solution $\bm{\beta}_0$ within $\mathcal{B}$. Thus,
$$
\inf_{\bm{\beta}: \|\bm{\beta}-\bm{\beta}_0\|\ge \varepsilon} \|\mathbf{M}(\bm{\beta})\| > 0 \quad\text{ for any }\varepsilon>0.
$$

We have verified the following conditions:
\begin{enumerate}[(i)]
  \item $\mathcal{B}$ is compact (Assumption~\ref{assum:compact});
  \item $\sup_{\bm{\beta}\in\mathcal{B}} \|\mathbf{H}_n(\bm{\beta}) - \mathbf{M}(\bm{\beta})\| = o_p(1)$;
  \item $\mathbf{M}(\bm{\beta}_0) = \mathbf{0}$, and $\inf_{\bm{\beta}: \|\bm{\beta}-\bm{\beta}_0\|\ge \varepsilon} \|\mathbf{M}(\bm{\beta})\| > 0$ for any $\varepsilon > 0$.
\end{enumerate}

These conditions correspond to the standard consistency conditions for Z-estimators \citep[Theorem 5.9]{van1998asymptotic}. Therefore,
$$
\hat{\bm{\beta}}_{\mathrm{FEM}} \xrightarrow{P} \bm{\beta}_0.
$$
This completes the proof.
\end{proof}

\medskip

\begin{proof}[Proof of Theorem \ref{thm:asymnormalFEM}]
By Theorem \ref{thm:consistencyFEM}, we already have $\hat{\bm{\beta}}_{\mathrm{FEM}} \xrightarrow{P} \bm{\beta}_0$.
Recall that the estimator solves the estimating equation
$$
\mathbf{H}_n(\hat{\bm{\beta}}_{\mathrm{FEM}}) = \mathbf{0},
$$
where
$$
\mathbf{H}_n(\bm{\beta}) = \frac{1}{n_0T} \sum_{j=1}^{T}
  \sum_{i \in \mathcal{I}_j} \left\{ y_i - b'(\mathbf{x}_{i}^{\top}\bm{\beta}+\hat{a}_j(\bm{\beta})) \right\}\mathbf{x}_{i}+\frac{a(\phi)}{2n_0T}\sum_{j=1}^{T} \frac{
  \sum_{i \in \mathcal{I}_j}
  b'''(\mathbf{x}_i^{\top}\bm{\beta} + \hat{a}_j(\bm{\beta})) \mathbf{x}_i
}{
  \sum_{i \in \mathcal{I}_j}
  b''(\mathbf{x}_i^{\top}\bm{\beta} + \hat{a}_j(\bm{\beta}))
}.
$$
By Taylor's expansion, we have
$$
\mathbf{H}_n(\hat{\bm{\beta}}_{\mathrm{FEM}})
= \mathbf{H}_n(\bm{\beta}_0)
+ \frac{\partial \mathbf{H}_n(\tilde{\bm{\beta}})}{\partial \bm{\beta}}
\big(\hat{\bm{\beta}}_{\mathrm{FEM}} - \bm{\beta}_0\big) = \mathbf{0},
$$
where $\tilde{\bm{\beta}}$ lies between $\hat{\bm{\beta}}_{\mathrm{FEM}}$ and $\bm{\beta}_0$. Because $\hat{\bm{\beta}}_{\mathrm{FEM}} \xrightarrow{P} \bm{\beta}_0$, we also have $\tilde{\bm{\beta}} \xrightarrow{P} \bm{\beta}_0$ as $n\to\infty$. Rearranging and scaling by $\sqrt{n}$ yields
$$
\sqrt{n}\bigl(\hat{\bm{\beta}}_{\mathrm{FEM}} - \bm{\beta}_0\bigr) =
- \Bigl( \frac{\partial \mathbf{H}_n(\tilde{\bm{\beta}})}{\partial \bm{\beta}} \Bigr)^{-1}
\sqrt{n}\, \mathbf{H}_n(\bm{\beta}_0).
$$
Differentiating $\mathbf{H}_n(\bm{\beta})$ with respect to $\bm{\beta}$ yields
$$
\frac{\partial\mathbf{H}_n}{\partial\bm{\beta}}
= -\frac{1}{n_0T}\sum_{j=1}^T\Bigl\{\sum_{i\in\mathcal{I}_j} b''(\mathbf{x}_{i}^{\top}\bm{\beta}+\hat{a}_j(\bm{\beta}))\mathbf{x}_i\mathbf{x}_i^\top-\frac{(\sum b''\mathbf{x}_i)(\sum b''\mathbf{x}_i)^\top}{\sum b''}\Bigr\}+\mathbf{R}(\bm{\beta}),
$$
where $\mathbf{R}(\bm{\beta}) = O(1/n_0)$.

For $n_0 \ge N_{0}$, the first term is symmetric negative definite (since $\bm{\Sigma}_{\mathbf{xx}}$ is positive definite) and the second term is dominated by the first term. Hence, $\partial \mathbf{H}_n(\bm{\beta})/\partial\bm{\beta}$ is uniformly negative definite on $\mathcal{B}$ and thus invertible.

Let
$$
\mathbf{J} = \lim_{T\to\infty} \frac{1}{T}\sum_{j=1}^T \Bigl(
\mathbb{E}\bigl[ b_{ij}'' \mathbf{x}_i \mathbf{x}_i^\top \bigr]
- \frac{ \mathbb{E}[ b_{ij}'' \mathbf{x}_i ] \mathbb{E}[ b_{ij}'' \mathbf{x}_i ]^\top }
       { \mathbb{E}[ b_{ij}'' ] }
\Bigr),
$$
where $b_{ij}'' = b''(\mathbf{x}_i^\top\bm{\beta}_0+a_{j,0})$ and expectations are taken with respect to the true distribution of observations within window $j$. Since the model is correctly specified and the covariate covariance matrix $\bm{\Sigma}_{\mathbf{xx}}$ is positive definite (Assumption~\ref{assum:x}), $\mathbf{J}$ is positive definite and invertible.

The uniform convergence of $\hat{a}_j(\bm{\beta})$ to $a_j^*(\bm{\beta})$ together with the consistency of
$\tilde{\bm{\beta}}$ implies
$$
\frac{\partial \mathbf{H}_n(\tilde{\bm{\beta}})}{\partial \bm{\beta}} \xrightarrow{P} -\mathbf{J}.
$$
For $1\leq j\leq T$, $\hat{a}_j(\bm{\beta}_0)$ is the unique solution of the following equation
\begin{equation}\label{eq:festep-real}
\frac{1}{n_0} \sum_{i \in \mathcal{I}_j}\left\{y_i - b'(\mathbf{x}_{i}^{\top}\bm{\beta}_0+a_{j})\right\}+ \frac{a(\phi)}{2n_0}\frac{\sum_{i \in \mathcal{I}_j}b'''( \mathbf{x}_{i}^{\top}\bm{\beta}_0+a_{j})}{\sum_{i \in \mathcal{I}_j}b''(\mathbf{x}_{i}^{\top}\bm{\beta}_0+a_{j})}=0.
\end{equation}
Let $\hat{a}_j(\bm{\beta}_0) = a_{j,0} + \delta_j$, where $a_{j,0}$ is the true intercept in window $j$. Let
$$
S_j = \frac{1}{n_0} \sum_{i \in \mathcal{I}_j} \varepsilon_i, \quad
B_j = \frac{1}{n_0} \sum_{i \in \mathcal{I}_j} b_{ij}'', \quad
C_j = \frac{1}{n_0} \sum_{i \in \mathcal{I}_j} b_{ij}''',
$$
where
$$
\varepsilon_i = y_i - b'(\mathbf{x}_i^{\top}\bm{\beta}_0+a_{j,0}), \quad b_{ij}'' = b''(\mathbf{x}_i^{\top}\bm{\beta}_0+a_{j,0}), \quad b_{ij}''' = b'''(\mathbf{x}_i^{\top}\bm{\beta}_0+a_{j,0}).
$$
Inserting $\hat{a}_j(\bm{\beta}_0) = a_{j,0} + \delta_j$ into Equation \eqref{eq:festep-real}, expanding $b'()$ to third order and the Firth correction term to first order in $\delta_j$, and collecting terms, we obtain

$$
S_j - B_j \delta_j - \frac{1}{2} C_j \delta_j^2 + \frac{a(\phi)}{2n_0} \frac{C_j}{B_j} + O_p(n_0^{-3/2})+ O(1/n_0^2)= 0.
$$
Solving this equation by a standard perturbation expansion yields
$$
\delta_j = \frac{S_j}{B_j} + O_p(1/n_0) + O(1/n_0^2),
$$
where the $O_p(1/n_0)$ term collects all zero-mean stochastic fluctuations, and the $O(1/n_0^2)$ term collects all higher-order deterministic bias components.

Note that
$$
\mathbf{H}_n(\bm{\beta}_0) = \frac{1}{n_0T} \sum_{j=1}^{T}
  \sum_{i \in \mathcal{I}_j} \left\{ y_i - b'(\mathbf{x}_{i}^{\top}\bm{\beta}_0+\hat{a}_j(\bm{\beta}_0)) \right\}\mathbf{x}_{i}+\frac{a(\phi)}{2n_0T}\sum_{j=1}^{T} \frac{
  \sum_{i \in \mathcal{I}_j}
  b'''(\mathbf{x}_i^{\top}\bm{\beta}_0 + \hat{a}_j(\bm{\beta}_0)) \mathbf{x}_i
}{
  \sum_{i \in \mathcal{I}_j}
  b''(\mathbf{x}_i^{\top}\bm{\beta}_0 + \hat{a}_j(\bm{\beta}_0))
}.
$$
Inserting $\hat{a}_j(\bm{\beta}_0) = a_{j,0} + \delta_j$ into $b'(\mathbf{x}_i^{\top}\bm{\beta}_0+\hat{a}_j(\bm{\beta}_0))$ and performing second-order Taylor expansion in $\delta_j$, we have

\begin{align}\label{eq:h1}
&\frac{1}{n_0T} \sum_{j=1}^{T}
  \sum_{i \in \mathcal{I}_j} \left\{ y_i - b'(\mathbf{x}_{i}^{\top}\bm{\beta}_0+\hat{a}_j(\bm{\beta}_0)) \right\}\mathbf{x}_{i} \notag\\
=& \frac{1}{n_0T} \sum_{j=1}^{T}
  \sum_{i \in \mathcal{I}_j}\Bigl\{\varepsilon_i - b_{ij}''\delta_j-\frac{1}{2} b_{ij}'''\delta_j^2+ O_p(\delta_j^3) \Bigr\}\mathbf{x}_{i} \notag\\
=& \frac{1}{n_0T} \sum_{j=1}^{T}
  \sum_{i \in \mathcal{I}_j}\Bigl\{\varepsilon_i - b_{ij}''\frac{S_j}{B_j}-\frac{1}{2} b_{ij}'''\frac{a(\phi)}{n_0B_j} \Bigr\}\mathbf{x}_{i}+O_p\Bigl(\frac1{n_0\sqrt{T}}\Bigr)+O\Bigl(\frac{1}{n_0^2}\Bigr) \notag\\
=& \frac{1}{n_0T} \sum_{j=1}^{T}
  \sum_{i \in \mathcal{I}_j}\Bigl\{\varepsilon_i - b_{ij}''\frac{S_j}{B_j}\Bigr\}\mathbf{x}_{i}-\frac{a(\phi)}{2n_0T}\sum_{j=1}^{T}\frac{
  \sum_{i \in \mathcal{I}_j}
 b_{ij}''' \mathbf{x}_i
}{
  \sum_{i \in \mathcal{I}_j}
  b_{ij}''
}+O_p\Bigl(\frac1{n_0\sqrt{T}}\Bigr)+O\Bigl(\frac{1}{n_0^2}\Bigr).
\end{align}
Inserting $\hat{a}_j(\bm{\beta}_0) = a_{j,0} + \delta_j$ into the Firth correction term  of $\mathbf{H}_n(\bm{\beta}_0)$,  and performing second-order Taylor expansion in $\delta_j$, we have
\begin{align}\label{eq:h2}
&\frac{a(\phi)}{2n_0T}\sum_{j=1}^{T} \frac{
  \sum_{i \in \mathcal{I}_j}
  b'''(\mathbf{x}_i^{\top}\bm{\beta}_0 + \hat{a}_j(\bm{\beta}_0)) \mathbf{x}_i
}{
  \sum_{i \in \mathcal{I}_j}
  b''(\mathbf{x}_i^{\top}\bm{\beta}_0 + \hat{a}_j(\bm{\beta}_0))
}  \notag\\
= &\frac{a(\phi)}{2n_0T}\sum_{j=1}^{T} \Bigl\{ \frac{
  \sum_{i \in \mathcal{I}_j}
 b_{ij}''' \mathbf{x}_i
}{
  \sum_{i \in \mathcal{I}_j}
  b_{ij}''
}
+\frac{(\sum b''''\mathbf{x}_i)(\sum b'')
-(\sum b''' \mathbf{x}_i)(\sum b''')}
{(\sum b'')^{2}}\delta_j+O_p(\delta_j^2)\Bigr\} \notag\\
=&\frac{a(\phi)}{2n_0T}\sum_{j=1}^{T}\frac{
  \sum_{i \in \mathcal{I}_j}
 b_{ij}''' \mathbf{x}_i
}{
  \sum_{i \in \mathcal{I}_j}
  b_{ij}''
}+O_p\Bigl(\frac1{n_0^{3/2}\sqrt{T}}\Bigr)+O\Bigl(\frac{1}{n_0^2}\Bigr).
\end{align}
Substituting \eqref{eq:h1} and \eqref{eq:h2} into $\mathbf{H}_n(\bm{\beta}_0)$ gives
\begin{align*}
\mathbf{H}_n(\bm{\beta}_0) &= \frac{1}{n_0T} \sum_{j=1}^{T}
  \sum_{i \in \mathcal{I}_j}\Bigl\{\varepsilon_i - b_{ij}''\frac{S_j}{B_j}\Bigr\}\mathbf{x}_{i}
+ O_p\Bigl(\frac1{n_0\sqrt{T}}\Bigr)+O\Bigl(\frac{1}{n_0^2}\Bigr)\\
&= \frac{1}{T} \sum_{j=1}^{T}
 \Bigl\{ \frac{1}{n_0} \sum_{i \in \mathcal{I}_j}
\varepsilon_i \mathbf{x}_i -  \frac{\frac{1}{n_0} \sum_{i \in \mathcal{I}_j} b_{ij}'' \mathbf{x}_i}{\frac{1}{n_0}\sum_{i \in \mathcal{I}_j} b_{ij}''}S_j\Bigr\}
+ O_p\Bigl(\frac1{n_0\sqrt{T}}\Bigr)+O\Bigl(\frac{1}{n_0^2}\Bigr)\\
&= \frac{1}{T} \sum_{j=1}^{T}
 \Bigl\{ \frac{1}{n_0} \sum_{i \in \mathcal{I}_j}
\varepsilon_i \mathbf{x}_i -  \Bigl(\frac{ \mathbb{E}[ b_{ij}'' \mathbf{x}_i ]}{\mathbb{E}[ b_{ij}'' ]}+O_p\Bigl(\frac{1}{\sqrt{n_0}}\Bigr)\Bigr)S_j\Bigr\}
+ O_p\Bigl(\frac1{n_0\sqrt{T}}\Bigr)+O\Bigl(\frac{1}{n_0^2}\Bigr)\\
&= \frac{1}{T} \sum_{j=1}^{T}
 \Bigl\{ \frac{1}{n_0} \sum_{i \in \mathcal{I}_j}
\varepsilon_i \mathbf{x}_i -  \frac{ \mathbb{E}[ b_{ij}'' \mathbf{x}_i ]}{\mathbb{E}[ b_{ij}'' ]}\frac{1}{n_0} \sum_{i \in \mathcal{I}_j} \varepsilon_i\Bigr\}
+ O_p\Bigl(\frac1{n_0\sqrt{T}}\Bigr)+O\Bigl(\frac{1}{n_0^2}\Bigr).
\end{align*}
Under $n_0/T^{1/3}\to\infty$, we have
\begin{align*}
\sqrt{n}\mathbf{H}_n(\bm{\beta}_0) &= \frac{1}{\sqrt{T}} \sum_{j=1}^{T}
\Bigl\{ \frac{1}{\sqrt{n_0}} \sum_{i \in \mathcal{I}_j}
\varepsilon_i \mathbf{x}_i -  \frac{ \mathbb{E}[ b_{ij}'' \mathbf{x}_i ]}{\mathbb{E}[ b_{ij}'' ]}\frac{1}{\sqrt{n_0}} \sum_{i \in \mathcal{I}_j} \varepsilon_i\Bigr\} + O_p\Bigl(\frac1{\sqrt{n_0}}\Bigr)+O\Bigl(\frac{\sqrt{T}}{n_0^{3/2}}\Bigr)\\
&=\frac{1}{\sqrt{T}}\sum_{j=1}^T (U_j - V_j) + o_p(1),
\end{align*}
where
$$
U_j = \frac{1}{\sqrt{n_0}} \sum_{i \in \mathcal{I}_j}
\varepsilon_i \mathbf{x}_i
\qquad \text{ and }\qquad
V_j = \frac{ \mathbb{E}[ b_{ij}'' \mathbf{x}_i ]}{\mathbb{E}[ b_{ij}'' ]}\frac{1}{\sqrt{n_0}} \sum_{i \in \mathcal{I}_j} \varepsilon_i.
$$
Obviously,
$$
\mathbb{E}[U_j - V_j]=\mathbb{E}[U_j]-\mathbb{E}[V_j]=\mathbf{0}.
$$
Due to
$$
\mathrm{Var}(U_j)=a(\phi)\mathbb{E}\bigl[ b_{ij}'' \mathbf{x}_i \mathbf{x}_i^\top \bigr],
$$

$$
\mathrm{Var}(V_j)=a(\phi)\frac{ \mathbb{E}[ b_{ij}'' \mathbf{x}_i ] \mathbb{E}[ b_{ij}'' \mathbf{x}_i ]^\top }
       { \mathbb{E}[ b_{ij}'' ]},
$$

$$
\mathrm{Cov}(U_j,V_j)=a(\phi)\frac{ \mathbb{E}[ b_{ij}'' \mathbf{x}_i ] \mathbb{E}[ b_{ij}'' \mathbf{x}_i ]^\top }
       { \mathbb{E}[ b_{ij}'' ]},
$$
we have
$$
\mathrm{Var}(U_j - V_j)=a(\phi)\Bigl(\mathbb{E}\bigl[ b_{ij}'' \mathbf{x}_i \mathbf{x}_i^\top \bigr]
- \frac{ \mathbb{E}[ b_{ij}'' \mathbf{x}_i ] \mathbb{E}[ b_{ij}'' \mathbf{x}_i ]^\top }
       { \mathbb{E}[ b_{ij}'' ] }\Bigr).
$$
By the central limit theorem and Assumption \ref{assum:infinity}, we have
$$
\frac{1}{\sqrt{T}}\sum_{j=1}^T \big(U_j - V_j\big)\xrightarrow{d} \mathcal{N}(\mathbf{0}, a(\phi)\mathbf{J}).
$$
It follows that
$$
\sqrt{n}\mathbf{H}_n(\bm{\beta}_0) \xrightarrow{d} \mathcal{N}(\mathbf{0}, a(\phi)\mathbf{J}).
$$
By Slutsky's theorem, we obtain
$$
\sqrt{n}\bigl(\hat{\bm{\beta}}_{\mathrm{FEM}} - \bm{\beta}_0\bigr)\xrightarrow{d} \mathcal{N}(\mathbf{0}, a(\phi)\mathbf{J}^{-1}).
$$
This completes the proof.
\end{proof}

\medskip

\begin{proof}[Proof of Theorem \ref{thm:consistency_bounds}]
Since $b''(\cdot) > 0$, $b'(\cdot)$ is continuous and strictly increasing. For the $l$-th block, and $\bm{\beta}\in\mathcal{B}$, let $\hat{\alpha}_{l}(\bm{\beta})$ denote the unique solution of
$$
S_l(\alpha,\bm{\beta}) = \sum_{i=l}^{l+w-1} \left\{y_i - b'(\mathbf{x}_{i}^{\top}\bm{\beta}+\alpha)\right\}=0,
$$
and let $\alpha_l^*(\bm{\beta})$ denote the unique solution of
$$
\bar{S}_l(\alpha,\bm{\beta})=\sum_{i=l}^{l+w-1}\mathbb{E}\left[y_i - b'(\mathbf{x}_{i}^{\top}\bm{\beta}+\alpha)\right]=0.
$$
By construction, $S_l(\hat{\alpha}_{l},\hat{\bm{\beta}}_{\mathrm{FEM}})=0$, thus $\hat{\alpha}_{l}=\hat{\alpha}_{l}(\hat{\bm{\beta}}_{\mathrm{FEM}})$. Since $\{(\mathbf{x}_{i},y_{i})\}_{i=l}^{l+w-1}$ are independent and the variances of individual terms in $S_l(\alpha,\bm{\beta})$ are uniformly bounded, the weak law of large numbers implies
$$
\frac{1}{w}S_l(\alpha,\bm{\beta}) \xrightarrow{P} \frac{1}{w}\bar{S}_l(\alpha,\bm{\beta}),\qquad \text{ as } w\to\infty.
$$
Since $b'(\cdot)$ is strictly increasing, both $S_l(\alpha,\bm{\beta})$ and $\bar{S}_l(\alpha,\bm{\beta})$ are strictly decreasing in $\alpha$. By the monotone-root argument, it follows that
$$
\hat{\alpha}_{l}(\bm{\beta}) \xrightarrow{P} \alpha_l^*(\bm{\beta}).
$$
Since $\hat{\alpha}_{l}(\bm{\beta})$ is stochastically equicontinuous, together with pointwise convergence, we have
$$
\sup_{\bm{\beta} \in \mathcal{B}}\left|\hat{\alpha}_{l}(\bm{\beta}) - \alpha_l^*(\bm{\beta}) \right| = o_p(1).
$$
By Theorem \ref{thm:consistencyFEM}, $\hat{\bm{\beta}}_{\mathrm{FEM}} \xrightarrow{P} \bm{\beta}_0$ and $\hat{\bm{\beta}}_{\mathrm{FEM}} \in \mathcal{B}$ with probability tending to one. Thus,
$$
\hat{\alpha}_{l}(\hat{\bm{\beta}}_{\mathrm{FEM}}) \xrightarrow{P} \alpha_l^*(\hat{\bm{\beta}}_{\mathrm{FEM}}).
$$
By the continuous mapping theorem,
$$
\alpha_l^*(\hat{\bm{\beta}}_{\mathrm{FEM}}) \xrightarrow{P} \alpha_l^*(\bm{\beta}_{0}).
$$
Let $\alpha_l^*=\alpha_l^*(\bm{\beta}_{0})$. Note that
$$
\hat{\alpha}_{l} - \alpha_l^* = (\hat{\alpha}_{l}(\hat{\bm{\beta}}_{\mathrm{FEM}})-\alpha_l^*(\hat{\bm{\beta}}_{\mathrm{FEM}}))+(\alpha_l^*(\hat{\bm{\beta}}_{\mathrm{FEM}})-\alpha_l^*),
$$
it follows that
$$
\hat{\alpha}_{l} \xrightarrow{P} \alpha_l^*.
$$
Since $w \le \min_{1 \le j \le k} n_j$, each block $B_{l}$ falls into one of two cases:
\begin{enumerate}[(i)]
  \item The block is fully contained in a single homogeneous sample, in which case $\alpha_l^*= c_j$ for some $j$;

  \item The block spans two adjacent samples $j$ and $j+1$, in which case the monotonicity of $b'(\cdot)$ implies that $\alpha_l^*$ is strictly bounded between $c_j$ and $c_{j+1}$.
\end{enumerate}
Moreover, for each $j$, there exists at least one block $B_{l_{j}}$ entirely lying within the $j$-th sample with true intercept $c_j$. Therefore,
$$
\overline{\beta}_0=\max_{1\leq j\leq k}c_{j}= \max\limits_{1\leq l\leq L}\alpha_l^*, \qquad
\qquad \underline{\beta}_0=\min_{1\leq j\leq k} c_{j}=\min_{1\leq l\leq L} \alpha_l^*.
$$
By the mean value theorem applied to $S_l(\cdot, \hat{\bm{\beta}}_{\mathrm{FEM}})$, there exists $\tilde{\alpha}_{l}$ between $\hat{\alpha}_{l}$ and $\alpha_l^*$ such that
$$
0 = S_l(\hat{\alpha}_{l},\hat{\bm{\beta}}_{\mathrm{FEM}})
= S_l(\alpha_l^*,\hat{\bm{\beta}}_{\mathrm{FEM}})
+ \frac{\partial S_l}{\partial \alpha}(\tilde{\alpha}_l,\hat{\bm{\beta}}_{\mathrm{FEM}})(\hat{\alpha}_{l} - \alpha_l^*).
$$
Rearranging the terms, we have
$$
\hat{\alpha}_{l} - \alpha_l^*
= -\Bigl( \frac{\partial S_l}{\partial \alpha}\big(\tilde{\alpha}_l,\hat{\bm{\beta}}_{\mathrm{FEM}}\big) \Bigr)^{-1}
S_l(\alpha_l^*,\hat{\bm{\beta}}_{\mathrm{FEM}}).
$$
Define
$$
\Psi_l(\alpha,\bm{\beta}) = \frac{1}{w}\frac{\partial S_l}{\partial \alpha}(\alpha,\bm{\beta})
= -\frac{1}{w}\sum_{i=l}^{l+w-1} b''(\mathbf{x}_i^{\top}\bm{\beta}+\alpha).
$$
Recall that for each $l$, $\alpha_l^*$ satisfies $\underline{\beta}_0 \le \alpha_l^* \le \overline{\beta}_0$. Let $\mathcal{K}=[\underline{\beta}_0 - 1, \overline{\beta}_0 + 1] \times \mathcal{B}$. Consider the function class
$$
\mathcal{F} = \{ \mathbf{x} \mapsto b''(\mathbf{x}^{\top}\bm{\beta}+\alpha) : (\alpha,\bm{\beta}) \in \mathcal{K} \}.
$$
Since $b''(\cdot)$ is continuous, $b''(\mathbf{x}^{\top}\bm{\beta}+\alpha)$ is continuous for every $\mathbf{x}$. Since $\mathbf{x}$ is sub-Gaussian, the envelope function $\sup_{(\alpha,\bm{\beta})\in\mathcal{K}} |b''(\mathbf{x}^\top\bm{\beta}+\alpha)|$ is integrable.  According to \citet[Example 19.8]{van1998asymptotic}, $\mathcal{F}$ is $P$-Glivenko-Cantelli. Therefore,
$$
\sup_{(\alpha,\bm{\beta}) \in \mathcal{K}} | \Psi_l(\alpha,\bm{\beta}) - \Psi^*(\alpha,\bm{\beta}) |
= o_p(1),
$$
where $\Psi^*(\alpha,\bm{\beta}) = -\mathbb{E}[ b''(\mathbf{x}^{\top}\bm{\beta}+\alpha) ]$.

By the triangle inequality,
$$
| \Psi_l(\tilde{\alpha} _l, \hat{\bm{\beta}}_{\mathrm{FEM}}) - \Psi^*(\alpha_l^*, \bm{\beta}_0) |\leq | \Psi_l(\tilde{\alpha} _l, \hat{\bm{\beta}}_{\mathrm{FEM}})
        - \Psi^*(\tilde{\alpha} _l, \hat{\bm{\beta}}_{\mathrm{FEM}}) | + | \Psi^*(\tilde{\alpha} _l, \hat{\bm{\beta}}_{\mathrm{FEM}})
            - \Psi^*(\alpha_l^*, \bm{\beta}_0)|.
$$
Since $\tilde{\alpha}_l$ is between $\hat{\alpha}_{l}$ and $\alpha_l^*$, $\hat{\alpha}_{l} \xrightarrow{P} \alpha_l^*$ implies $\tilde{\alpha}_l \xrightarrow{P} \alpha_l^*$. Together with
$\hat{\bm{\beta}}_{\mathrm{FEM}} \xrightarrow{P} \bm{\beta}_0$, $(\tilde{\alpha}_l, \hat{\bm{\beta}}_{\mathrm{FEM}}) \in \mathcal{K}$ with probability tending to one. The first term is bounded by the uniform convergence bound and thus is $o_p(1)$. By the dominated convergence theorem, $\Psi^*(\alpha,\bm{\beta})$ is continuous. Since $(\tilde{\alpha}_l, \hat{\bm{\beta}}_{\mathrm{FEM}}) \xrightarrow{P} (\alpha_l^*, \bm{\beta}_0)$, by the continuous mapping theorem, the second term is also $o_p(1)$. Therefore,
$$
\Psi_l(\tilde{\alpha}_l, \hat{\bm{\beta}}_{\mathrm{FEM}}) \xrightarrow{P} \Psi^*(\alpha_l^*, \bm{\beta}_0).
$$
Let
$$
c_{\min} = \frac{1}{2} \min_{\alpha \in [\underline{\beta}_0, \overline{\beta}_0]}
\mathbb{E}[ b''(\mathbf{x}^{\top}\bm{\beta}_0+\alpha) ] > 0.
$$
With probability tending to one,
$$
\frac{1}{w}\frac{\partial S_l}{\partial \alpha}(\tilde{\alpha}_l, \hat{\bm{\beta}}_{\mathrm{FEM}})
= \Psi_l(\tilde{\alpha}_l, \hat{\bm{\beta}}_{\mathrm{FEM}}) \leq -c_{\min},
$$
which is equivalent to
$$
\frac{\partial S_l}{\partial \alpha}(\tilde{\alpha}_l, \hat{\bm{\beta}}_{\mathrm{FEM}}) \leq -w c_{\min}.
$$
It follows that
$$
|\hat{\alpha}_{l} - \alpha_l^*|
\le \frac{1}{w c_{\min}} | S_l(\alpha_l^*,\hat{\bm{\beta}}_{\mathrm{FEM}}) |
$$
with probability tending to one.

Since $\bar{S}_l(\alpha_l^*,\bm{\beta}_0)=0$, we have
$$
\frac{1}{w}S_l(\alpha_l^*,\hat{\bm{\beta}}_{\mathrm{FEM}})
= \underbrace{\frac{1}{w}[S_l(\alpha_l^*,\hat{\bm{\beta}}_{\mathrm{FEM}}) - S_l(\alpha_l^*,\bm{\beta}_0)]}_{\textstyle :=\, \delta_{1,l}}
+ \underbrace{\frac{1}{w}[S_l(\alpha_l^*,\bm{\beta}_0) - \bar{S}_l(\alpha_l^*,\bm{\beta}_0)]}_{\textstyle :=\, \delta_{2,l}}.
$$
By the mean value theorem,
\begin{align*}
\frac{1}{w} [S_l(\alpha_l^*,\hat{\bm{\beta}}_{\mathrm{FEM}}) -  S_l(\alpha_l^*,\bm{\beta}_0)]
&=\frac{1}{w}\sum_{i=l}^{l+w-1} \left\{b'(\mathbf{x}_{i}^{\top}\bm{\beta}_0+\alpha_l^*)-b'(\mathbf{x}_{i}^{\top}\hat{\bm{\beta}}_{\mathrm{FEM}}+\alpha_l^*)\right\}\\
&=\frac{1}{w}\sum_{i=l}^{l+w-1} b''(\mathbf{x}_{i}^{\top}\tilde{\bm{\beta}}_{l,i}+\alpha_l^*)\mathbf{x}_{i}^{\top}(\bm{\beta}_0-\hat{\bm{\beta}}_{\mathrm{FEM}}),
\end{align*}
where $\tilde{\bm{\beta}}_{l,i}$ is between $\hat{\bm{\beta}}_{\mathrm{FEM}}$ and $\bm{\beta}_0$.

Since $\hat{\bm{\beta}}_{\mathrm{FEM}} \xrightarrow{P} \bm{\beta}_0$, $(\alpha_l^*, \tilde{\bm{\beta}}_{l,i}) \in \mathcal{K}$ with probability tending to one. Since $\mathbf{x}$ is sub-Gaussian, the random variables $b''(\mathbf{x}_{i}^{\top}\tilde{\bm{\beta}}_{l,i}+\alpha_l^*)\|\mathbf{x}_i\|$ satisfy the Bernstein moment condition uniformly over $\mathcal{K}$. Applying Bernstein's inequality and a union bound over the $L$ overlapping blocks gives
$$
\max_{1 \leq l \leq L}\frac{1}{w}\sum_{i=l}^{l+w-1} b''(\mathbf{x}_{i}^{\top}\tilde{\bm{\beta}}_{l,i}+\alpha_l^*)\|\mathbf{x}_i\| = O_p(1).
$$
Thus,
$$
\max_{1 \leq l \leq L} |\delta_{1,l}| \leq \|\hat{\bm{\beta}}_{\mathrm{FEM}} - \bm{\beta}_0\| \cdot \max_{1 \leq l \leq L}\frac{1}{w}\sum_{i=l}^{l+w-1} b''(\mathbf{x}_{i}^{\top}\tilde{\bm{\beta}}_{l,i}+\alpha_l^*)\|\mathbf{x}_i\| = o_p(1)\cdot O_p(1) = o_p(1).
$$
For any fixed $(\alpha,\bm{\beta})\in \mathcal{K}$, define
$$
\xi_i = y_i - b'(\mathbf{x}_i^\top \bm{\beta}+\alpha)
- \mathbb{E}\!\Bigl[y_i - b'(\mathbf{x}_i^\top \bm{\beta}+\alpha)\Bigr].
$$
The terms $\{\xi_i\}_{i=l}^{l+w-1}$ are independent and have zero mean. Since $y_i$ belongs to an exponential family, its conditional distribution given $\mathbf{x}_i$ has a finite moment generating function in a neighborhood of zero. Together with the sub-Gaussian assumption on $\mathbf{x}_i$, this yields the Bernstein moment condition. That is, there exist constants $\sigma^2, M > 0$ independent of $(\alpha,\bm{\beta})$ such that
$$
\mathbb{E}[|\xi_i|^k] \leq \frac{1}{2} k! \,\sigma^2 M^{k-2},\quad \forall\,k\ge 2.
$$
By Bernstein's inequality,
\[
P\left( \left| \frac{1}{w} S_l(\alpha,\bm{\beta}) - \frac{1}{w} \bar{S}_l(\alpha,\bm{\beta}) \right| > t \right)
\leq 2 \exp\left( -\frac{w t^2}{2(\sigma^2 + M t)} \right).
\]
For sufficiently small $t > 0$ satisfying $t \leq \sigma^2 / M$, we have $\sigma^2 + M t \leq 2\sigma^2$. This yields the sub-Gaussian-type tail bound
\[
P\left( \left| \frac{1}{w} S_l(\alpha,\bm{\beta}) - \frac{1}{w} \bar{S}_l(\alpha,\bm{\beta}) \right| > t \right)
\leq 2\exp\left(-C w t^2\right),
\]
where $C = 1/(4\sigma^2) > 0$.

For each block $B_{l}$ and sufficiently small $t > 0$ satisfying $t \leq \sigma^2 / M$,
$$
P\left( |\delta_{2,l}| > t \right)=P\left( \left| \frac{1}{w} S_l(\alpha_l^*,\bm{\beta}_0) - \frac{1}{w} \bar{S}_l(\alpha_l^*,\bm{\beta}_0) \right| > t \right) \le  2\exp\left(-C w t^2\right).
$$
Applying a union bound over all $L = n - w + 1$ blocks yields
\[
P\left( \max_{1 \le l \le L} |\delta_{2,l}| > t \right)\le \sum_{l=1}^{L} P\left( |\delta_{2,l}| > t \right)\le 2L \exp\left(-C w t^2\right)< 2n \exp\left(-C w t^2\right).
\]
By Assumption \ref{assum:w}, $w / \log n \to \infty$ as $n\to\infty$. Thus, for any fixed $t > 0$ satisfying $t \leq \sigma^2 / M$,
\[
n \exp\left(-C w t^2\right) = \exp\left( \log n - C w t^2 \right) \to 0,
\]
which implies
\[
\max_{1 \le l \le L} |\delta_{2,l}| = o_p(1).
\]
Consequently,
$$
\max_{1 \leq l \leq L} |\hat{\alpha}_{l} - \alpha_l^*|\le\frac{1}{c_{\min}} \max_{1 \leq l \leq L} |\delta_{1,l} + \delta_{2,l}| = o_p(1).
$$
Since $\hat{\overline{\beta}}_0=\max\limits_{1\leq l\leq L}\hat{\alpha}_{l}$ and $\overline{\beta}_0= \max\limits_{1\leq l\leq L}\alpha_l^*$ , we have
$$
|\hat{\overline{\beta}}_0 - \overline{\beta}_0| = | \max_{1 \leq l \leq L} \hat{\alpha}_{l} - \max_{1 \leq l \leq L} \alpha_l^* | \leq \max_{1 \leq l \leq L} |\hat{\alpha}_{l} - \alpha_l^*| = o_p(1).
$$
Similarly, we have
$$
|\hat{\underline{\beta}}_0 - \underline{\beta}_0| = | \min_{1 \leq l \leq L} \hat{\alpha}_{l} - \min_{1 \leq l \leq L} \alpha_l^* | \leq \max_{1 \leq l \leq L} |\hat{\alpha}_{l} - \alpha_l^*| = o_p(1).
$$
Therefore,
$$
\hat{\overline{\beta}}_0 \xrightarrow{P} \overline{\beta}_0 \quad \text{and} \quad \hat{\underline{\beta}}_0 \xrightarrow{P} \underline{\beta}_0\quad \text{ as } n \to \infty.
$$
This completes the proof.
\end{proof}

\newpage
\bibliographystyle{plainnat}
\bibliography{GLM_reference}

\end{document}